\documentclass[12pt,preprint]{emulateapj}
\def \msun{\hbox{${\rm M}_\odot$}}
\def \smass{\hbox{$\rm M_*/M_{\odot}$}}
\def \fred{\hbox{$f_{red}$}}
\def \fredT{\hbox{$f_{rT}$}}
\def \rgrp{\hbox{$r_{grp}$}}
\def \SigmaF{\hbox{$\Sigma_5$}}
\def \r200{\hbox{$R_{200}$}}
\def \bgc{\hbox{$B_{gc}$}}
\def \MSgrp{\hbox{M$_{*,grp}$}}
\def \SMgrp{\hbox{M$_{*,grp}$}}
\def\gtapr {\lower .1ex\hbox{\rlap{\raise .6ex\hbox{\hskip .3ex
        {\ifmmode{\scriptscriptstyle >}\else
                {$\scriptscriptstyle >$}\fi}}}
        \kern -.4ex{\ifmmode{\scriptscriptstyle \sim}\else
                {$\scriptscriptstyle\sim$}\fi}}}
\def\ltapr {\lower .1ex\hbox{\rlap{\raise .6ex\hbox{\hskip .3ex
        {\ifmmode{\scriptscriptstyle <}\else
                {$\scriptscriptstyle <$}\fi}}}
        \kern -.4ex{\ifmmode{\scriptscriptstyle \sim}\else
                {$\scriptscriptstyle\sim$}\fi}}}

\shortauthors{I.H. Li et al.}
\shorttitle{Galaxy Groups in the RCS-1 Sample}

\begin{document}

\title{Evolution of Group Galaxies from the First Red-Sequence Cluster Survey
}
\author{I.H. Li\altaffilmark{1,2}}
\email{tli@astro.swin.edu.au}
\author{H.K.C. Yee\altaffilmark{2}}
\email{hyee@astro.utoronto.ca}
\author{B.C. Hsieh \altaffilmark{3}}
\email{bchsieh@asiaa.sinica.edu.tw}
\author{M. Gladders \altaffilmark{4,5}}
\email{gladders@oddjob.uchicago.edu}
\altaffiltext{1}{Centre for Astrophysics \& Supercomputing, Swinburne University of Technology, PO Box 218, Hawthorn, Victoria 3122, Australia}
\altaffiltext{2}{50 St. George Street, Department of Astronomy \& Astrophysics, University of Toronto, Toronto, ON, Canada, M5S 3H4}
\altaffiltext{3}{Institute of Astronomy and Astrophysics, Academia Sinica, PO Box 23-141, Taipei 106, R.O.C. Taiwan}
\altaffiltext{4}{Department of Astronomy and Astrophysics, University of Chicago, 5640 S. Ellis Ave, Chicago, IL 60637}
\altaffiltext{5}{Kavli Institute for Cosmological Physics, University of Chicago, 5640 South Ellis Avenue, Chicago, IL 60637}


\begin{abstract}
We study the evolution of the red galaxy fraction ($f_{red}$) in 905 galaxy 
groups with $0.15 \leq z < 0.52$. 
The galaxy groups are identified by the `probability Friends-of-Friends' algorithm from the first Red-Sequence Cluster Survey (RCS1) photometric-redshift sample.
There is a high degree of uniformity in the properties of the red-sequence 
of the group galaxies, indicating that the luminous red-sequence galaxies 
in the groups are already in place by $z\sim0.5$ and that they have a formation
epoch of $z\gtapr2$.
In general, groups at lower redshifts exhibit larger $f_{red}$ than those at higher redshifts, showing a group Butcher-Oemler effect.
We investigate the evolution of \fred~by examining its dependence on four
parameters, which can be classified as one intrinsic and three environmental: galaxy stellar mass (M$_*$), total group stellar mass
(\SMgrp, a proxy for group halo mass), normalized group-centric radius
(\rgrp), and local galaxy density (\SigmaF). 
We find that M$_*$ is the dominant parameter such that there is a strong
correlation between \fred~and galaxy stellar mass. 
Furthermore, the dependence
of \fred~on the environmental parameters is also a strong function of M$_*$.
Massive galaxies (M$_*\gtapr 10^{11}$ \msun) show little dependence of \fred~on
\rgrp, \SMgrp, and \SigmaF~over the redshift range.
The dependence of \fred~on these parameters is primarily seen for
galaxies with lower masses, especially for M$_*\ltapr 10^{10.6}$ \msun.
We observe an apparent `group down-sizing' effect, in that
galaxies in lower-mass halos, after controlling for galaxy
stellar mass, have lower \fred. 
We find a dependence of \fred~on both \rgrp~and \SigmaF~after the other
parameters are controlled.
At a fixed \rgrp, there is a significant dependence of \fred~on \SigmaF,
while \rgrp~gradients of \fred~are seen for galaxies in similar \SigmaF~regions.
This indicates that galaxy group environment has a residual effect over that
of local galaxy density (or vice versa), and both parameters need to
be considered.
This result suggests that processes identified with local galaxy density,
such as galaxy harassment and mergers, and those associated with
accretion into a larger group halo, such as ram pressure and strangulation,
are both partaking in driving galaxies to their final red quiescent state.
We discuss these results in the context of the `nature vs nurture'
scenario of galaxy evolution.
\end{abstract}

\section{Introduction} \label{introduction}
	Galaxies evolve with time. Generally speaking, the star formation rate within galaxies decreases as galaxies age, and hence galaxy colors transit from blue to red, and their spiral arms become less and less dominant. It has been known since the 1970's that red galaxies tend to populate galaxy clusters and blue galaxies are common in the field \citep{1974ApJ...194....1O}.
The fraction of red galaxies decreases from cluster center to the field. Measuring galaxy morphologies in 55 clusters in the nearby Universe, \citet{1980ApJ...236..351D} found that galaxy population fractions have a stronger dependence on local galaxy density than on cluster-centric radius. 
This work formulated the well-known `morphology-density' relation, and is generally investigated as the `color-density' relation later on in the literature \citep[e.g.,][]{2001ApJ...562L...9K, 2006MNRAS.370..198C, 2007ApJ...654..138Q,2007ApJS..172..270C,2007MNRAS.376.1445C}.
This result is interpreted as a decrease in the star formation rate as local galaxy density increases. 
A `critical density' was suggested to characterize the local galaxy density effect \citep[e.g.,][]{2002MNRAS.334..673L,2003ApJ...584..210G,2004AJ....128.2677T} with the concept that star formation rate and galaxy color exhibit abrupt changes when the galaxy density crosses the critical density. 
Regardless of whether this critical density exists or what its true meaning is, the density where both star formation rates and galaxy colors change rapidly may indicate an environmental transition between galaxy groups and the field \citep[e.g.,][hereafter L09]{2001ApJ...562L...9K,2003PASJ...55..757G,2004MNRAS.348.1355B,2009ApJ...698...83L}.

	Galaxy groups are small galaxy aggregations bound by gravity. 
Some groups contain a few galaxies while others may have several dozens.
This gives galaxy groups a fundamental role in building up large-scale structures, as more and more galaxies are accreted into groups over
time and groups merge together. 
In the local Universe the majority of galaxies are found in galaxy groups \citep[e.g.,][]{1983ApJS...52...61G,1998ApJ...503..518F,2004MNRAS.348..866E}.
Because of their sufficiently high galaxy density and low velocity dispersion, galaxy groups are a favored environment for interactions and mergers, 
causing most of the galaxy transformations in morphology, star formation rates, and colors
\citep[e.g.,][]{1998MNRAS.300..146G, 1998ApJ...496...73M,2000ApJ...530..652H}.
The suppressed star formation in galaxy groups hence will lead to a decline in the mean star formation rate of the Universe as structures grow with decreasing redshift.
This scenario provides a plausible explanation for the declining global star formation rate with cosmic time \citep[e.g.,][]{1996MNRAS.283.1388M,1996ApJ...460L...1L,1999AJ....118..603C,2008MNRAS.383.1058C}.
	
	Within a galaxy group, the properties of galaxies exhibit a dependence on their distance from the group center. 
Observationally, galaxies in the center of groups are more likely to be bright, red, and early-type. 
The fraction of early-type galaxies decreases with increasing group-centric radius, while the fraction of fainter, bluer, and late-type galaxies has a positive correlation with group-centric radius \citep[e.g.,][]{2002MNRAS.335..825D,2006MNRAS.370.1223B,2006MNRAS.366....2W}.
Galaxies in the centers of low-mass groups already show different properties from those in the field \citep[e.g.,][]{2007MNRAS.374.1169B,2008ApJ...672L.103K}. 
Such differences are stronger in higher-mass groups, and galaxies in these groups have properties close to those in clusters \citep[e.g.,][]{2002MNRAS.335..825D}. 
Galaxy groups, therefore, can be considered as `mini-clusters'. 
Yet galaxy groups are systems typically smaller than clusters by a factor of $\sim$10 or more in mass. 
Mechanisms such as ram-pressure stripping and harassment are perhaps not important within galaxy groups. 
This makes galaxy groups useful sites to study galaxy evolution 
possibly free of complicated multiple driving mechanisms.

Most studies on galaxy groups, however, are in the local Universe using samples from the Sloan Digital Sky Survey (SDSS) and the Two-degree-field (2dF) Survey \citep[e.g.,][]{2004MNRAS.348.1355B,2006MNRAS.370.1003M, 2006ApJ...642..188P,2006ApJ...652.1077R}.
At $z > 0.1$, our current understanding of galaxy groups is based on samples considerably
 smaller than that available in the local Universe; for examples, the CNOC2 \citep[e.g.,][]{2001ApJ...563..736C,2005MNRAS.358...71W,2005MNRAS.358...88W,2009MNRAS.398..754B}, DEEP2 \citep[e.g.,][]{2007MNRAS.376.1445C,2007MNRAS.376.1425G}, and zCOSMOS \citep[e.g.,][]{2009ApJ...697.1842K,2010A&A...509A..40I} group catalogs.
Galaxy groups are not easily detected beyond the local Universe because of their low gas content, as well as the less significant number-density contrast to the background. 
Traditionally, galaxy groups are identified within spectroscopic-redshift samples using algorithms such as the friends-of-friends or the Voronoi tessellation
 method \citep[e.g.,][]{2001ApJ...563..736C,2005ApJ...625....6G,2006ApJS..167....1B, 2006MNRAS.369.1334S,2008A&A...479..927T}.
The requirement of large spectroscopic surveys with high completeness adds difficulties to the study of galaxy groups at $z>0.1$, since obtaining such catalogs requires huge amounts of time, money, and work. 
Another route to identify galaxy groups is to apply dedicated group-finding algorithms to photometric-redshift samples \citep[e.g.,][]{2004MNRAS.349..425B,2008AJ....135..809L}.
Galaxy groups found this way have much larger uncertainties in their redshift estimation, and contain background galaxies as their members. 
The contamination of false groups may also bias scientific analyses. 
Even so, the ability to obtain large samples of photometric-redshift galaxy groups will still allow us to gain useful information,
after correcting for background contamination statistically.

	We study galaxy groups at $0.15 \leq z < 0.52$ using a photometric-redshift group sample drawn from the multi-band photometric data of the first Red-Sequence Cluster Survey \citep[RCS1;][]{2005ApJS..157....1G}. 
We aim to explore how colors of galaxies, which are a proxy for galaxy population, change with redshift, and how environment affects
 galaxy populations therein.
We follow the methodology used in our study of groups associated with CNOC1 clusters (L09) in deriving photometric redshifts, identifying galaxy groups, and defining environmental parameters.
The structure of this paper is as follows. 
We describe our data and group sample in \S \ref{sec:data} and \S \ref{sec:sample}. 
The results are presented in \S\ref{sec:results}, and discussed in \S\ref{sec:discussion}. We summarize our work in \S\ref{sec:summary}.
We adopt the cosmological parameters $H_0$=70 km/s/Mpc, $\Omega_m$=0.3, and $\Omega_{\Lambda}$=0.7. 

\section{The Data}\label{sec:data}
\subsection{The Observation}
We derive our galaxy group sample from the Northern patches of 
the RCS1, which is an imaging survey with the primary goal of 
measuring $\Omega_m$ and $\sigma_8$ using the cluster mass function \citep{2005ApJS..157....1G,2007ApJ...655..128G}. 
The survey covers 22 widely separated patches in total in the Northern and Southern sky with a total area of  92 square degrees. 
The observations were carried out from May 1999 to January 2001. 
The imaging in $R_c$ and $z'$ of the ten Northern patches was conducted using the CFH-12K camera on the 3.6m CFHT telescope. 
The camera has 12k$\times$8k pixels in total, consisting of twelve 2k$\times$4k CCDs with 0.206 $\arcsec$ per pixel, providing a $42\arcmin\times 28\arcmin$ field of view.
The layout of each patch is arranged with 15 pointings in a slightly overlapping grid of $3\times 5$ pointings, 
giving a patch size of $2.1\times 2.3$ deg$^2$. 
The integration times are 900 and 1200 seconds with average seeing of
$\sim 0.70\arcsec$ and $\sim 0.62\arcsec$ in the $R_c$ and $z'$ passbands, respectively. 
The $5\sigma$ limiting magnitudes for point sources are $R_c=24.8$ (Vega) and $z_{AB}=23.9$ in an aperture of diameter 2.7$\arcsec$. 
The details of the data reduction and photometric catalog can be found in 
Gladders \& Yee (2005).

To obtain photometric redshifts, we require additional photometric bands.
The $B$ and $V$ photometry for the RCS1 Northern patches was obtained
 as a follow-up project using the CFH-12K camera \citep{2005ApJS..158..161H}.
It covers 33.6 deg$^2$ in total with 108 pointings ($\sim 75\%$ of the original RCS1-CFHT patches). 
This defines the sky area used for our group catalog.
The runs were carried-out from May 2001 to June 2002. 
The typical exposure times in $B$ and $V$ are 840s and 480s, respectively. 
The average seeing is $\sim 0.95\arcsec$ in $B$ and $\sim 0.65\arcsec$ in $V$. 
The $5\sigma$ limiting magnitudes are $B$=25.0 and $V$=24.5, on average,
 within an aperture of 2.7$\arcsec$ diameter.
We refer to \cite{2005ApJS..158..161H} for more details on the follow-up data set.

\subsection{The Photometric-Redshift Catalog}
	We use the empirical photometric-redshift method modified from \cite{2008AJ....135..809L} to estimate galaxy redshifts. 
This photometric redshift algorithm assumes that galaxy redshift is a 
polynomial function of galaxy magnitudes and colors, and the coefficients are derived from a training set which is a catalog containing spectroscopic galaxy redshifts and multi-band photometry. 
We construct the redshift training set using spectroscopic samples
from the Hubble Deep Field \citep[HDF;][]{2004ApJ...600L..93G},
the Canadian Network for Observational Cosmology Survey \citep[CNOC2;][]{2000ApJS..129..475Y}, and the Deep Extragalactic Exploratory Probe \citep[DEEP2;][]{2005ApJS..159...41V}. 
The photometry of the training set is processed and calibrated in the same manner as the RCS1 catalogs, and the details can be found in \citet{2005ApJS..158..161H}. 
The final training set contains $\sim$5,300 galaxies with $BVRz'$ photometry.

Instead of dividing the training set and all input galaxies into 
several fixed color-magnitude cells as was done in \cite{2008AJ....135..809L},
we derive the coefficients of the photometric-redshift polynomial fit
individually for each input galaxy by using a subset of the training set
 which contains 400 galaxies chosen from the complete training set whose magnitudes and colors are the closest to the input  galaxy.
The choice of the size of the sub-training set is a compromise
between accuracy and computation time. Our tests show that any sub-training
set with more than 250 galaxies will provide reasonable photometric
redshift results.
These training galaxies are chosen based on quadratically summed 
ranks of color and magnitude differences between
the training set galaxies and the input galaxy.
All four magnitudes and six colors are used.
In doing the fitting, we further assign weights to the chosen 
training-set galaxies, based
on the inverse value of their final rank.
This partially alleviates the redshift
bias that may be introduced to input galaxies near the edges of 
the color-magnitude distributions.
The photometric-redshift errors are estimated empirically by assuming 
Gaussian magnitude errors and bootstrapping the training set in the color-magnitude cell.
We remove galaxies with very large photometric redshift uncertainties
from the sample.
A galaxy is considered not to have an acceptable photometric redshift if
its uncertainty is larger than the very loose criterion of 0.6(1+$z$).
As in L09, we assign a weight, $w_i$, to each galaxy  
based on the inverse of the fraction of galaxies with acceptable photometric redshifts to the total as a function of $R_c$ magnitude.
This $w_i$ is found not strongly dependent on galaxies colors \citep{2005ApJ...629L..77Y, 2009ApJ...698...83L}.
The effect of $w_i$ on our results will be further discussed in \S\ref{subsec:uncertainty}.

\begin{figure}
        \includegraphics[width=8.2cm]{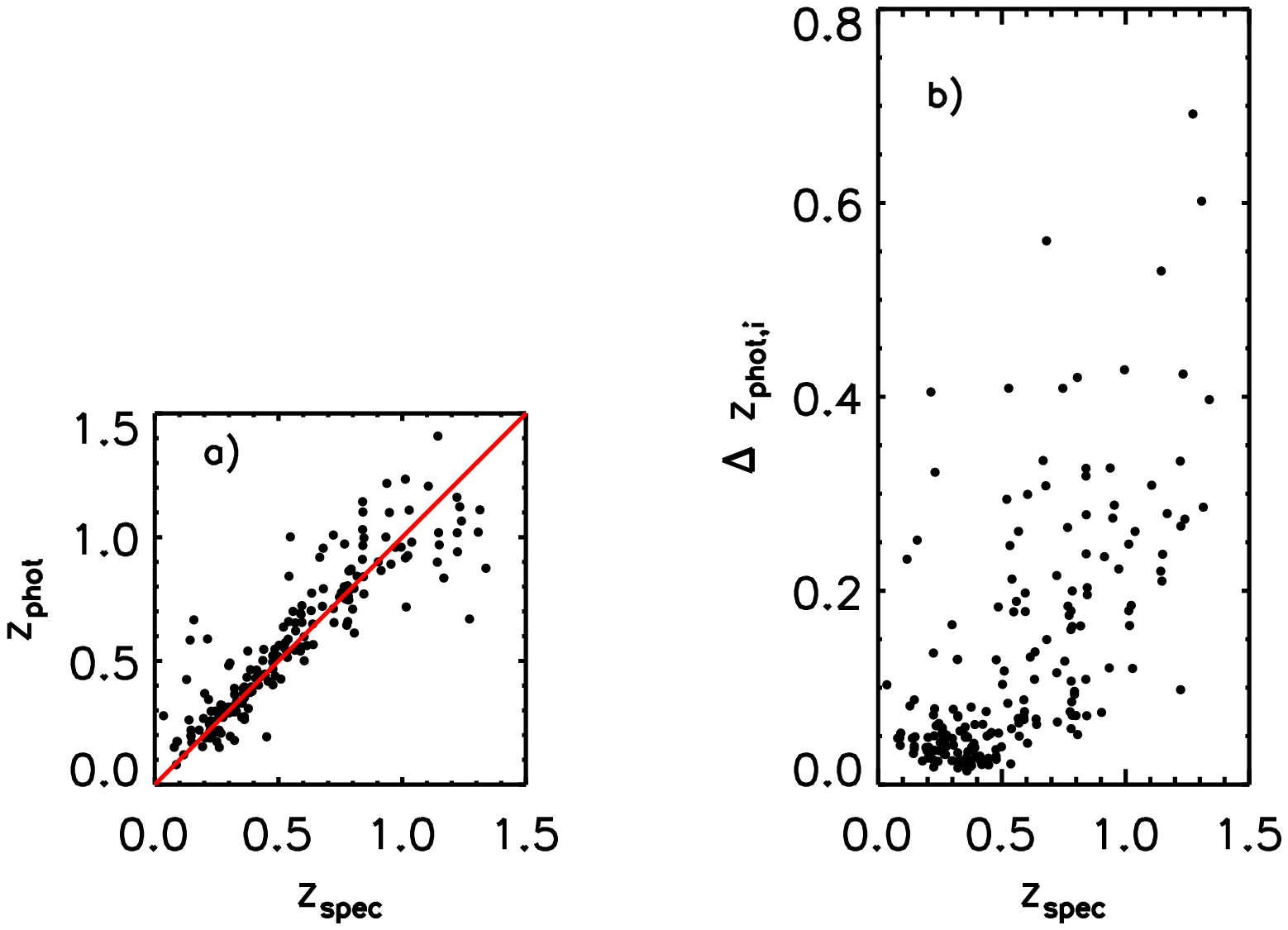}
        \includegraphics[width=8.2cm]{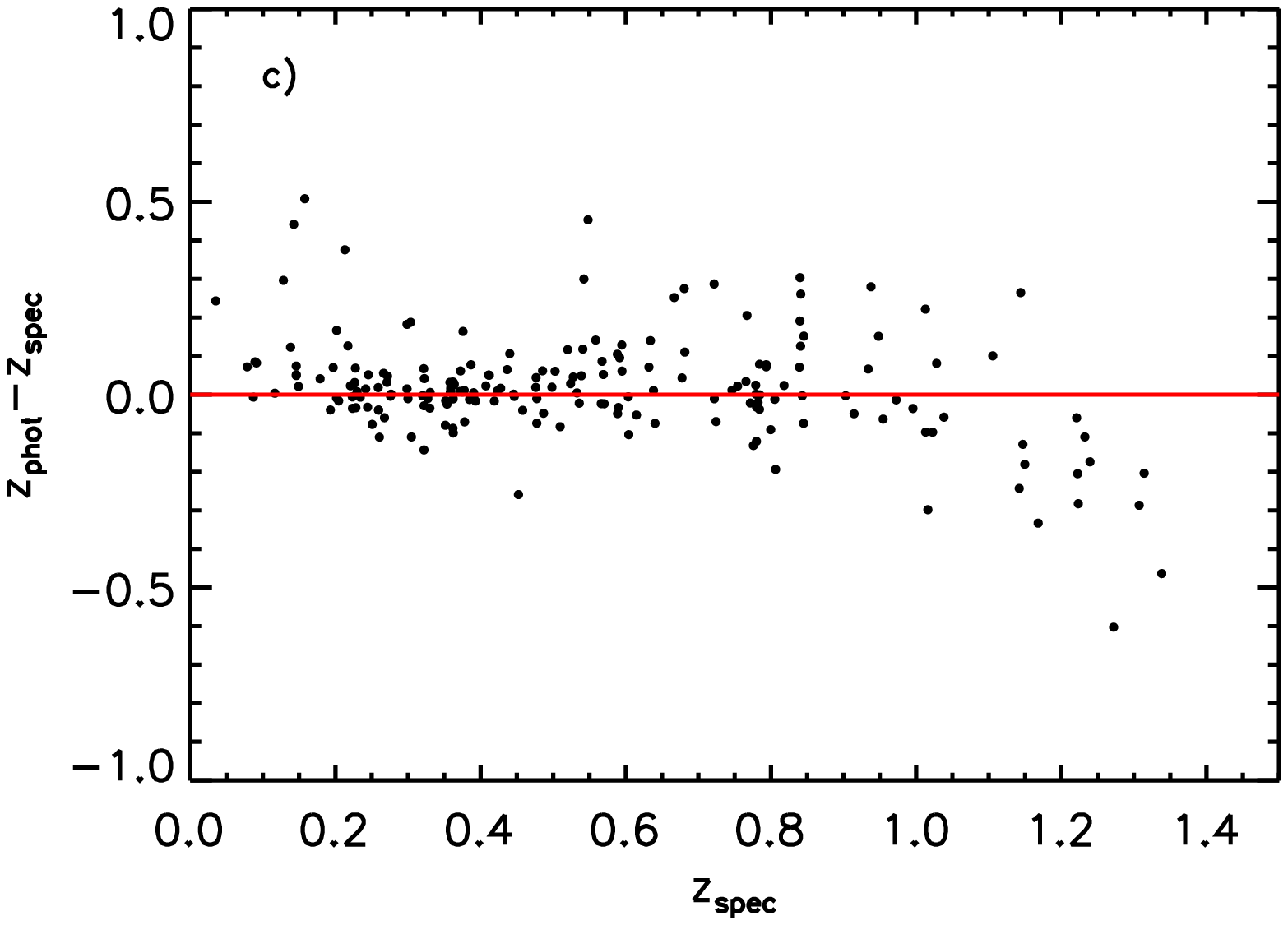}
        \caption{Panels a) and c):
Comparisons between photometric and spectroscopic redshifts for 200 galaxies whose photometric redshifts are derived using the rest of the $\sim$5100 training set galaxies.
The 1$\sigma$ dispersion of $z_{phot}-z_{spec}$ is $\sim$0.064 for $0.15\leq z \leq 0.6$.
Panel b): Empirically estimated 1-$\sigma$ photometric redshift uncertainties of individual galaxies as a function of $z_{spec}$.
\label{fig:photoz}}
        \end{figure}

To test the photometric-redshift accuracy, we randomly exclude 200 galaxies from the training set and treat them as input galaxies.
Their photometric redshifts are then derived using the rest of the training set galaxies, and compared with their known spectroscopic redshifts, as
presented in Figure \ref{fig:photoz}, along with their estimated uncertainties.
For galaxies within 
our redshift range of interest ($0.15 \leq z_{spec} \leq 0.6$),
the overall 1$\sigma$ dispersion (Figure \ref{fig:photoz}) is $\sim$0.064
with a mean $z_{phot}-z_{spec}$ off set of 0.011.
More tests of our photometric-redshift technique using both simulated data
and the GOODS-N catalog can be found in \citet{2010A&A...523A..31H}.

\section{The Galaxy Group Sample} \label{sec:sample}
\subsection{Group Catalogs and the Sample}
The identification of galaxy groups is achieved using the \it `probability Friends-of-Friends' (pFoF) \rm algorithm, described and tested in detail in \citet{2008AJ....135..809L}. 
The algorithm uses the group redshift probability density as a conditional 
probability density and the photometric-redshift probability distribution
of individual galaxies to perform the friends-of-friends linkage in the
third dimension.
A major feature of this algorithm is that the group redshift probability 
density is improved as more linked group members are added.

The pFoF algorithm has been tested using a simulated catalog \citep{2006MNRAS.365...11C} based on the Virgo Consortium Millennium Simulation \citep{2005Natur.435..629S}. 
Different linking criteria in both the 2D sky plane and the redshift direction, as well as various sample depths, have been adopted to investigate the group-finding performance. 
The results of these tests are described at length in \citet{2008AJ....135..809L}.
The RCS group catalogs are obtained by applying the pFoF algorithm to a sample of depth $M^*_{R_c}+2.0$, where $M^*_{R_c}$=-21.41 \citep{1997A&A...320...41K}.
We have generated a complete catalog of galaxy groups which is available in electronic form (Table \ref{grpsample}). 
Galaxy groups in this catalog are selected to have $N_{gal} \ge 5$ and $N_{gz} \ge 5$, and redshift between  $z$=0.15 and $z$=$z_{cut}$.
Here, $N_{gal}$ is the net weighted member count to $M^*_{R_c}+2.0$
(i.e., $N_{gal}=\sum(w_i)-N_{bg}$, where the computation of the background
counts $N_{bg}$ is presented in \S3.5); $N_{gz}$ is the actual number 
count of the linked galaxies in the pFoF group; and
$z_{cut}$ is the redshift where the nominal $R_c$ limit of a patch has 
a depth of $M^*_{R_c}+1.5$.
The $R_c$ limit is defined as the magnitude at which $w_i$ reaches 2.0, and ranges from 21.16 to 22.86 for the different patches, corresponding to a $z_{cut}$ range of 0.30 to 0.53.
The $N_{gal}$ and $N_{gz}$ limits of $\ge5$ are chosen as a compromise between
having a sample reaching down to sufficiently low group mass and the
number of false detections, based on the results of our mock tests. 

These criteria  produce a sample of 1153 groups.
Because we use a photometric redshift technique modified from that of
Li \& Yee (2008), we redo the mock tests for the pFoF algorithm.
The results, which are similar to those in Li \& Yee (2008),
 show that this sample may contain $\sim$35\% false groups 
(see \S\ref{subsec:uncertainty}).
 However, we can use other information to attempt to
cut down the fraction of false groups.  For examples, the total stellar
mass of a group \MSgrp~and the richness 
parameter $B_{gc}$ are measured using net $\rm {M}_*$ and galaxy counts,
respectively, within an aperture on the sky, rather than based on FoF
connected galaxy counts, and thus provide different information 
(see \S\ref{Bgc} for the calculations of \SMgrp~and $B_{gc}$).
Thus, groups with small \MSgrp~or $B_{gc}$, and in particular
groups with very discrepant \MSgrp and \bgc~(i.e., large \bgc
small \MSgrp, or vice versa) are expected to more likely be false groups.
This is verified by our mock group sample tests.
Applying the same \MSgrp~cut to the measured \MSgrp~in our mock catalog, we 
find that $\sim65$\% of the false groups have log(\SMgrp$/M_{\sun}$)$\leq$11.2.
For our analyses, we further employ a total group stellar mass cut of 
log(\SMgrp$/M_{\sun}$)$\geq$11.2 and a richness cut of $B_{gc}  \geq 125$
Mpc$^{1.8}h_{50}^{-1.8}$ in selecting our group sample. 
The total group stellar mass and richness cuts remove 248 groups 
from the sample.
Assuming that these are mostly false detections,
we expect $\sim$15\% false groups in the remaining group catalog, which
we use as our final group sample.
We note that the most significant effect that false groups produce in our
analysis is to bias the galaxy red fraction towards smaller values 
(see \S\ref{subsec:uncertainty} and \citet{2008AJ....135..809L}),
especially for samples of  poor groups where the false detection rate is higher.

\begin{figure}
        \includegraphics[angle=90,width=8.2cm]{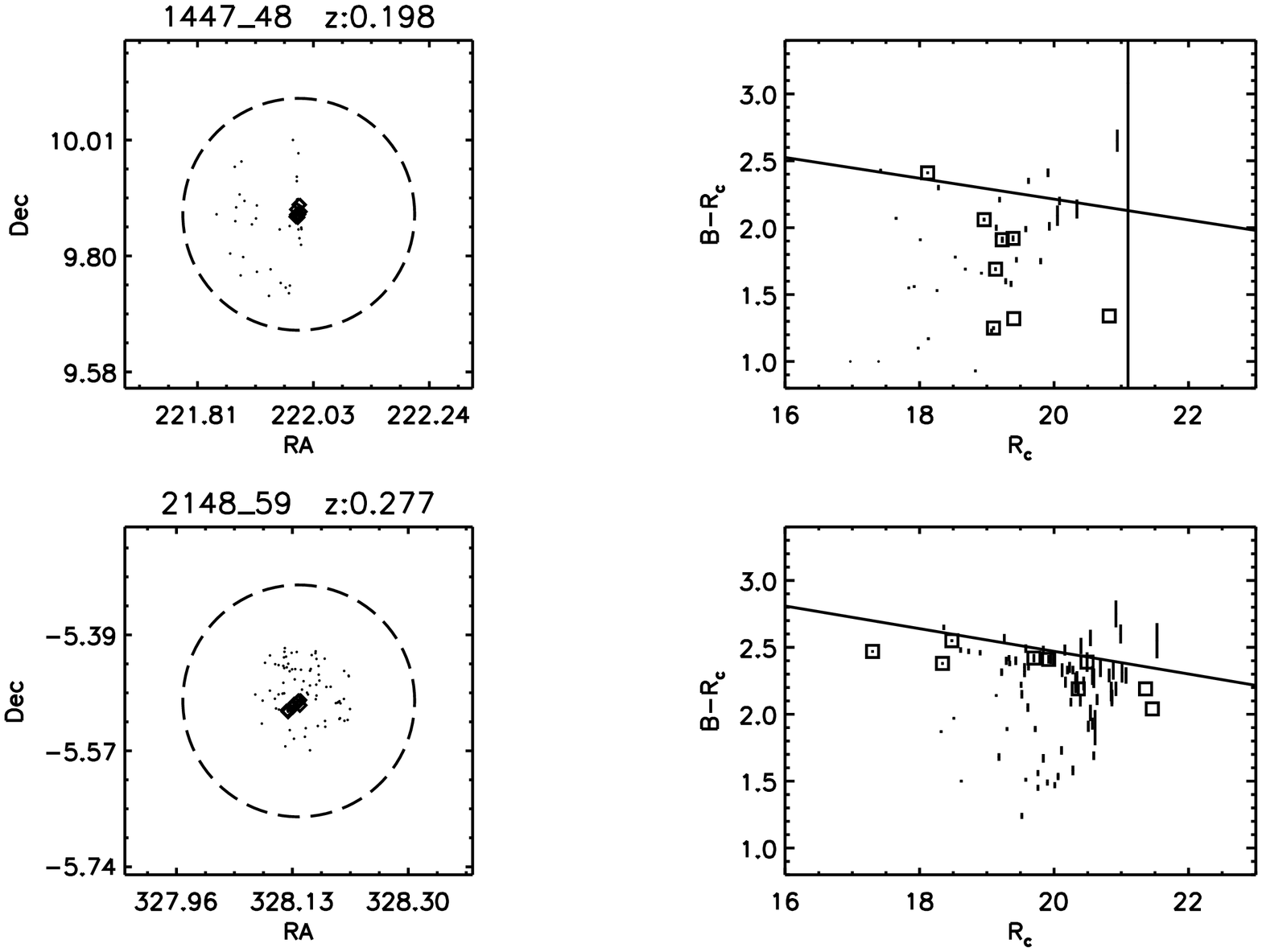}
        \includegraphics[angle=90,width=8.2cm]{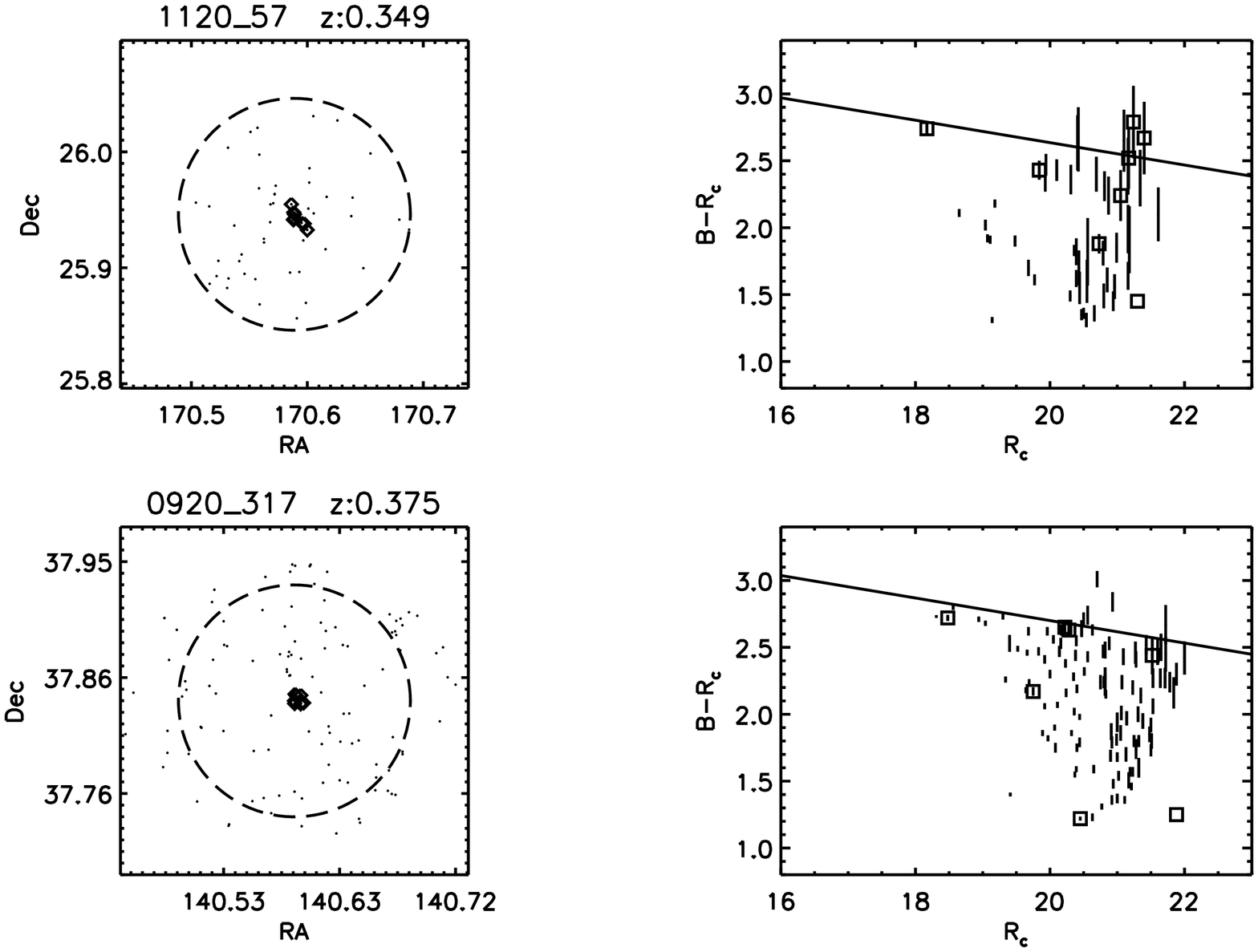}
        \includegraphics[angle=90,width=8.2cm]{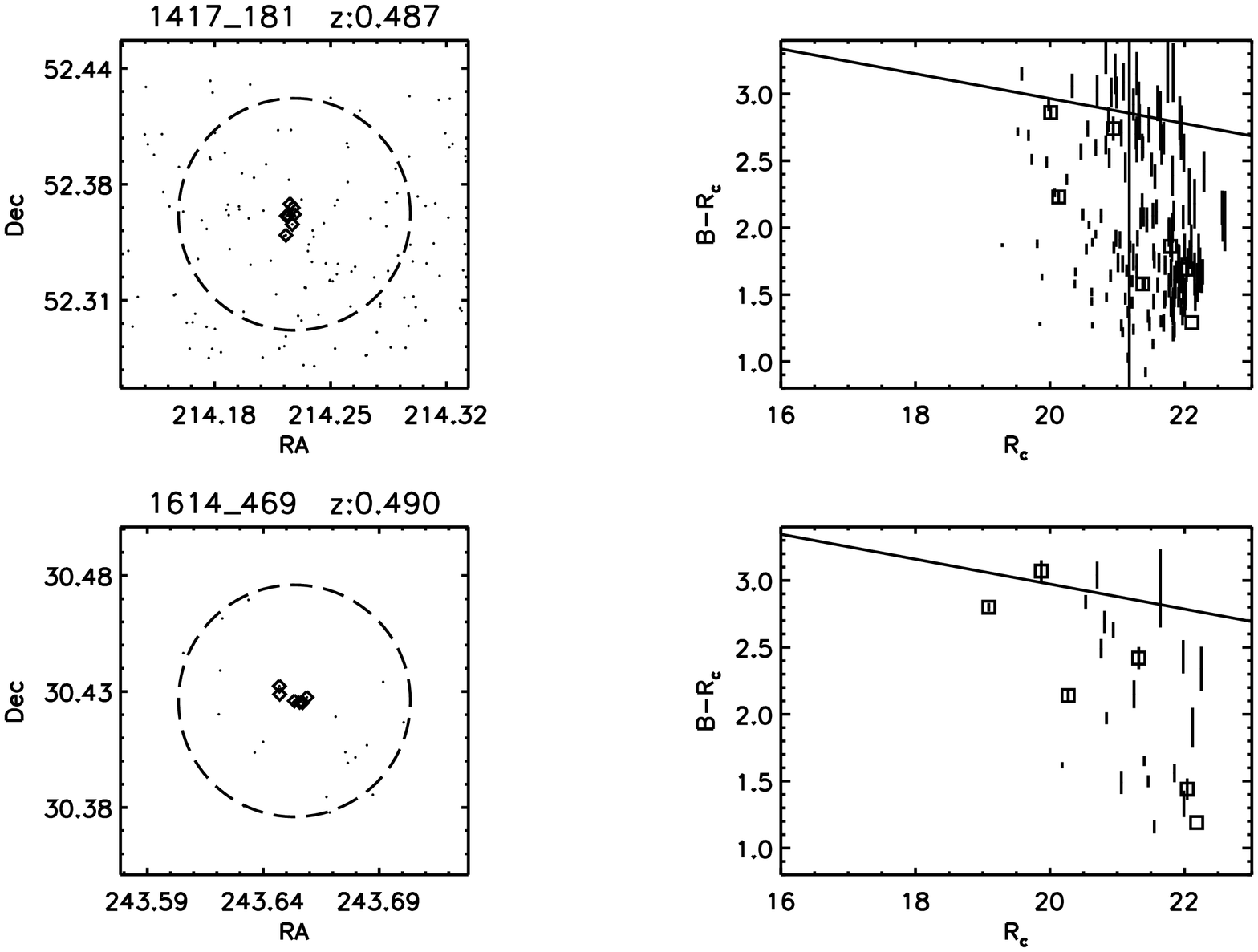}
        \includegraphics[angle=90,width=8.2cm]{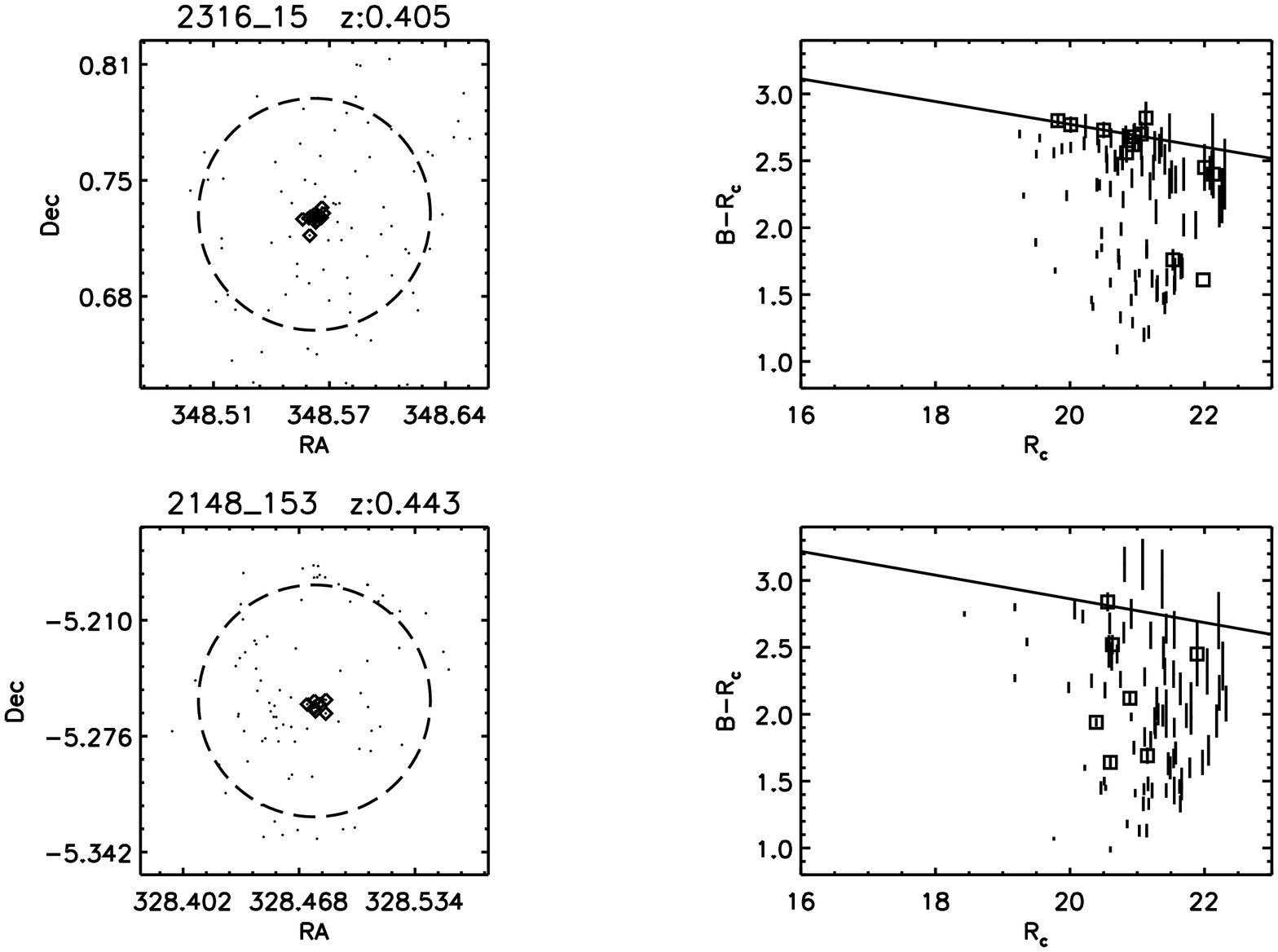}
        \caption{Examples of sky locations and observed $B-R_c$ versus $R_c$
color-magnitude diagrams for RCS galaxy groups.
The group ID and redshift are indicated in the title of each plot.
The circles in the sky maps have a radius of one $R_{200}$.
The linked pFoF members are plotted by squares, while other objects are plotted as dots.
The solid lines in the color-magnitude diagrams are theoretical red sequences based on group redshifts.
\label{tmap}}
        \end{figure}

\begin{figure}
        \includegraphics[angle=90,width=8.2cm]{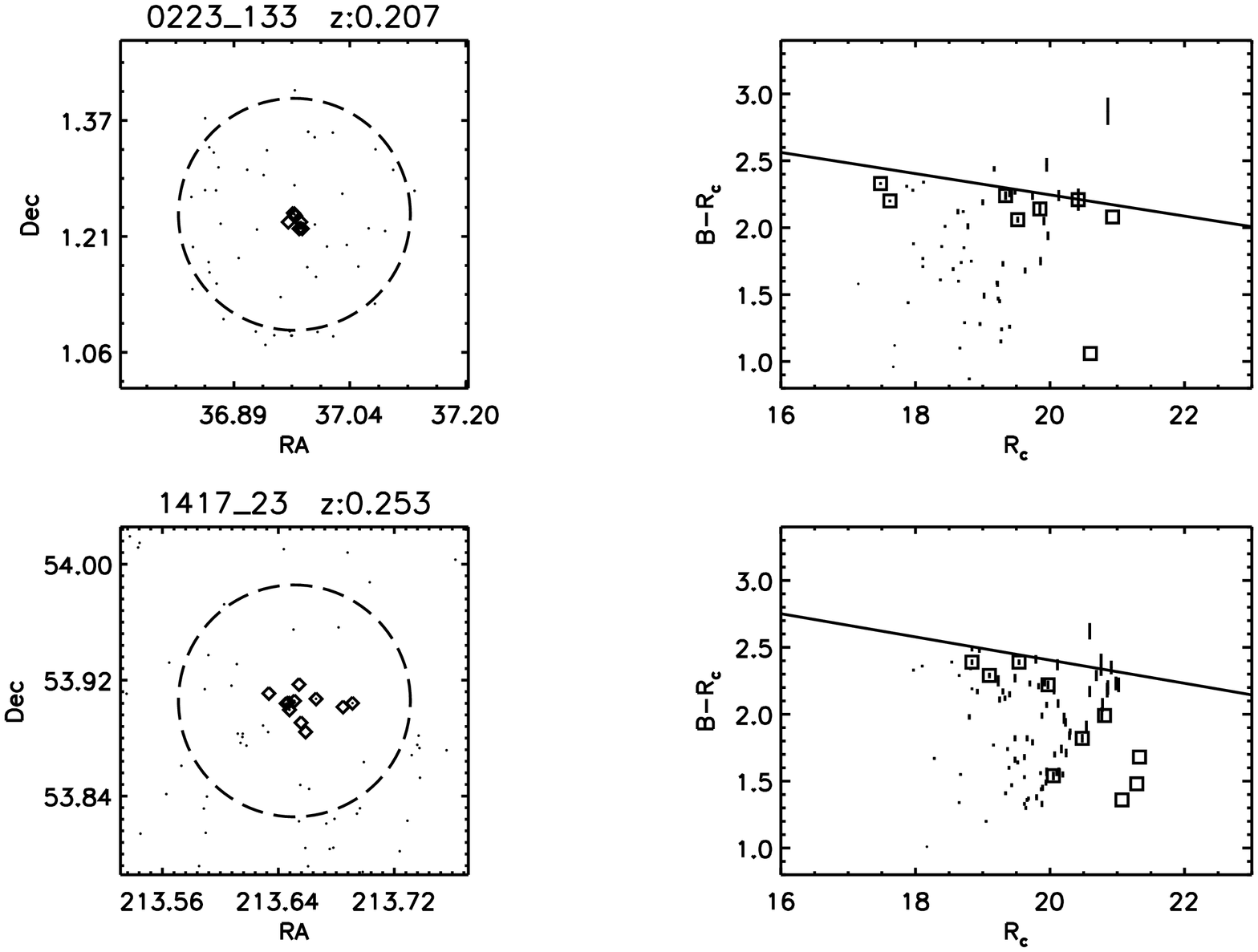}
        \includegraphics[angle=90,width=8.2cm]{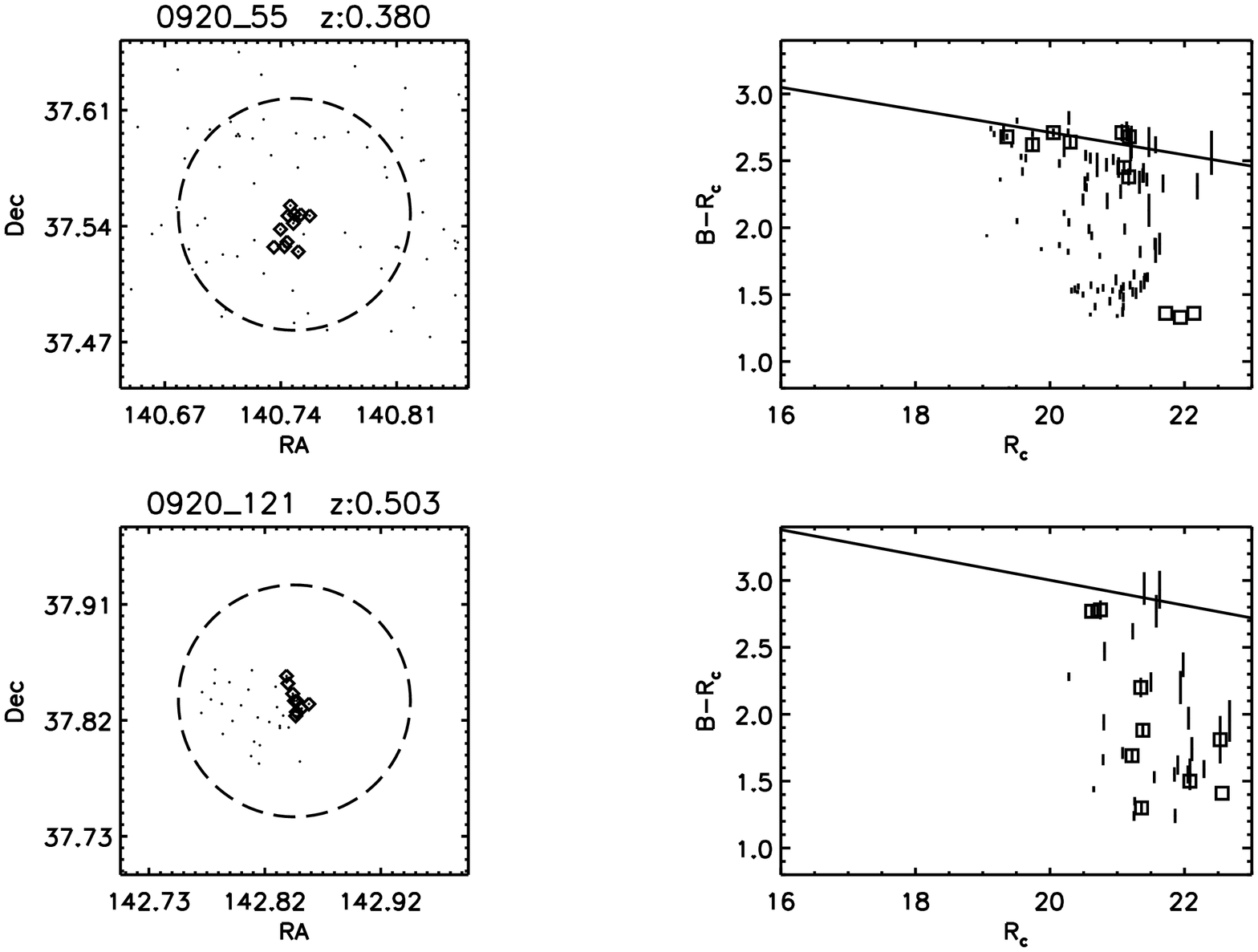}
        \includegraphics[angle=90,width=8.2cm]{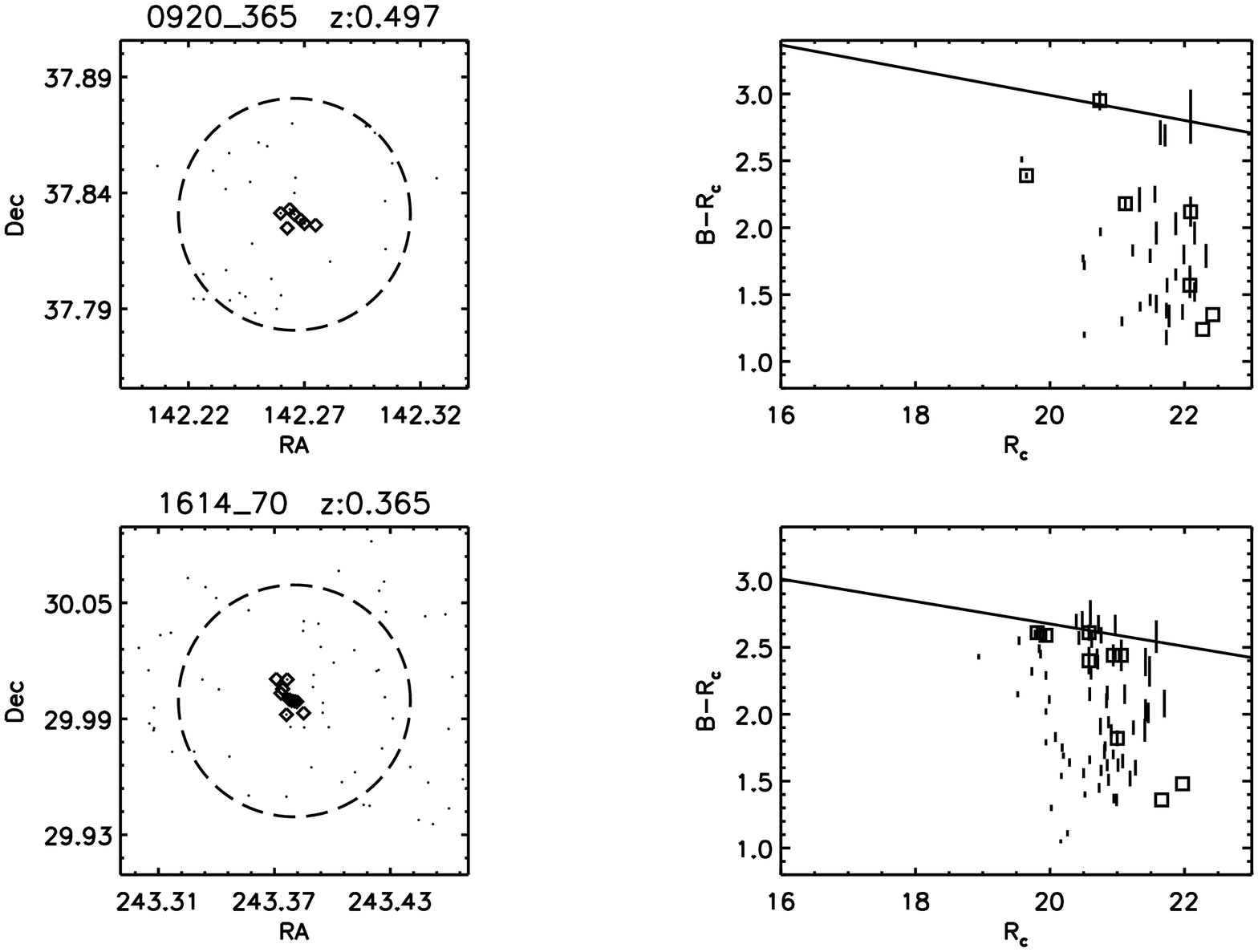}
        \includegraphics[angle=90,width=8.2cm]{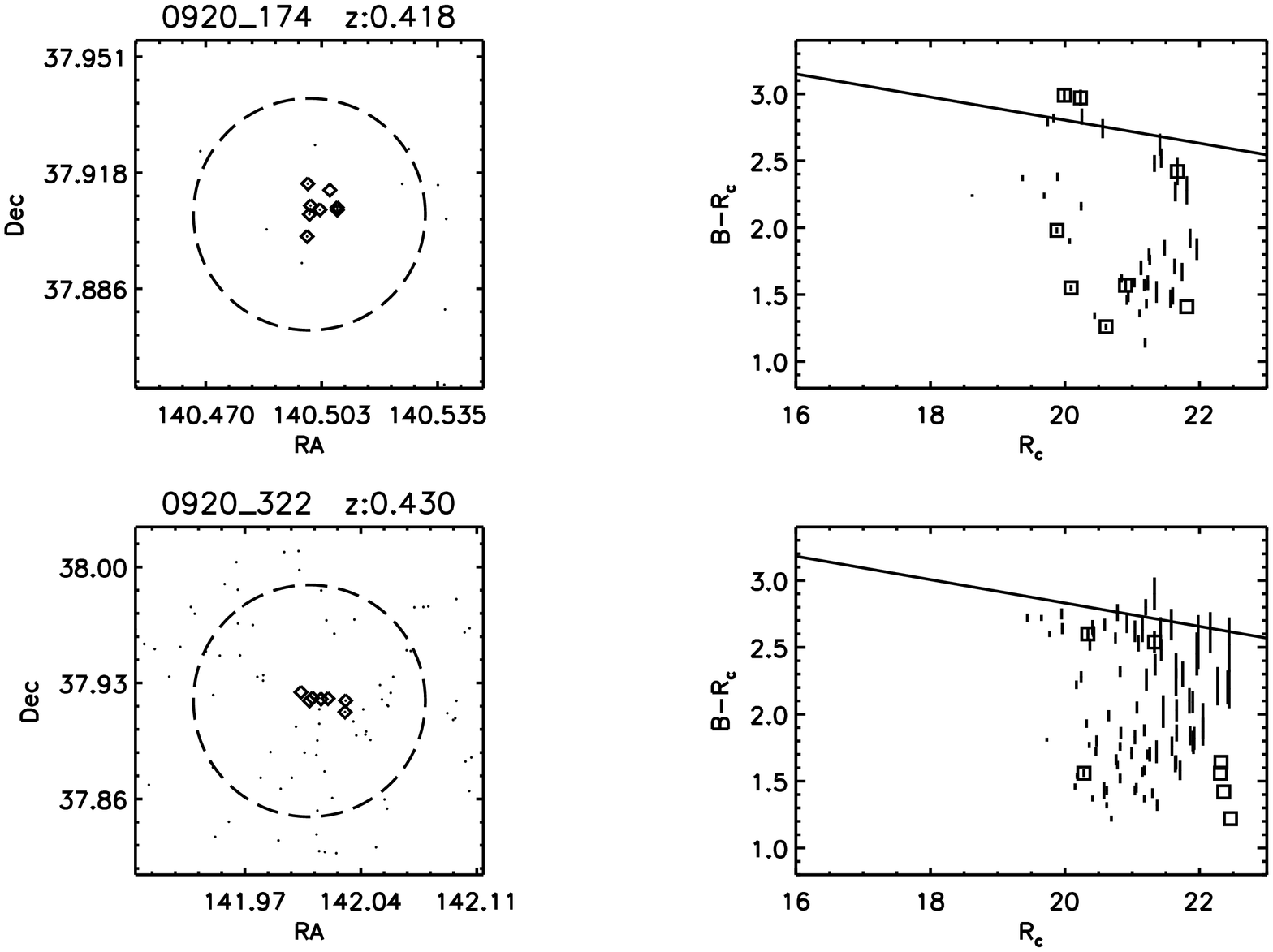}
        \caption{Same as Figure \ref{tmap}. \label{tmap2}}
        \end{figure}

\begin{figure}
        \includegraphics[angle=90,width=8.2cm]{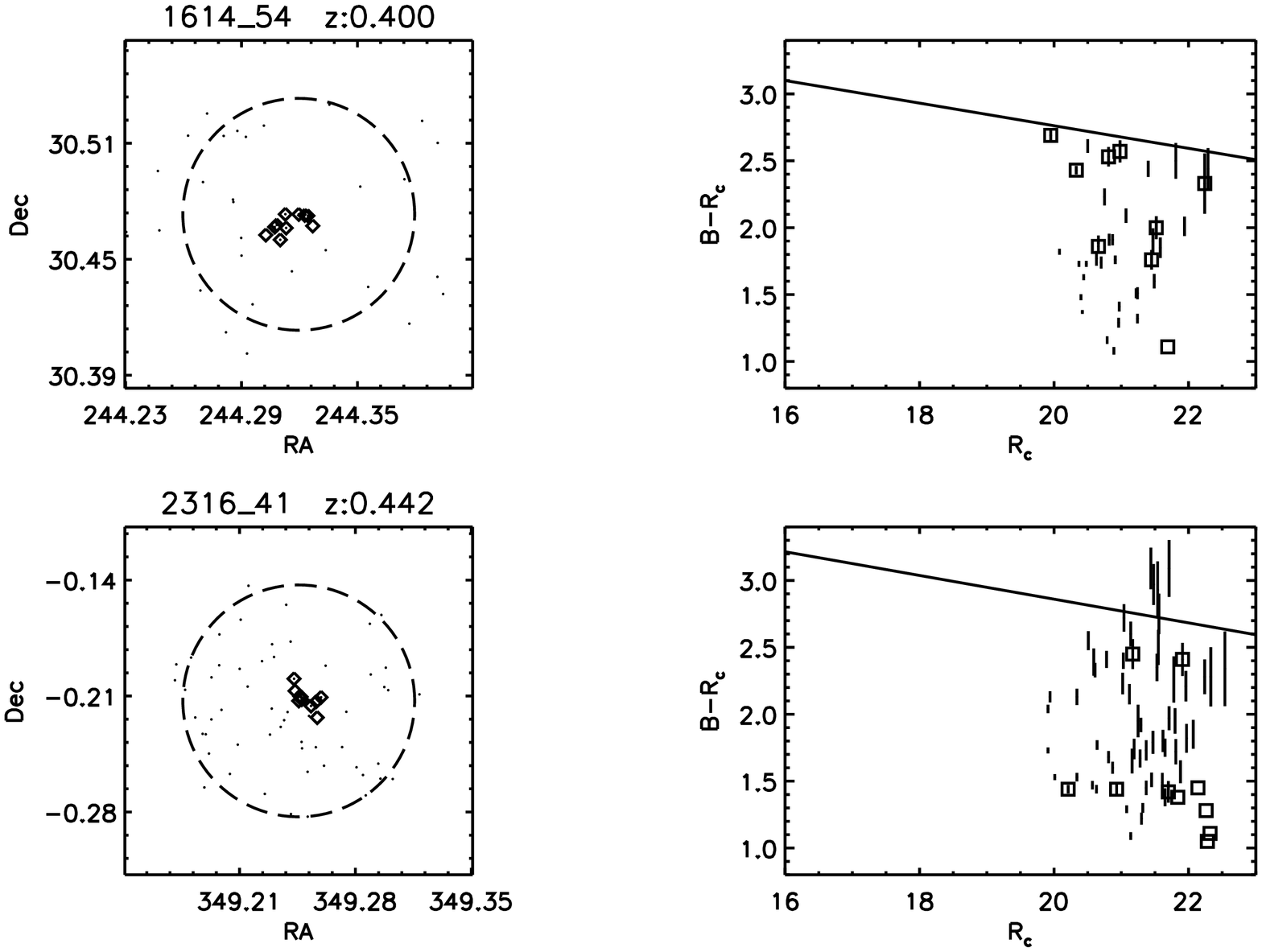}
        \includegraphics[angle=90,width=8.2cm]{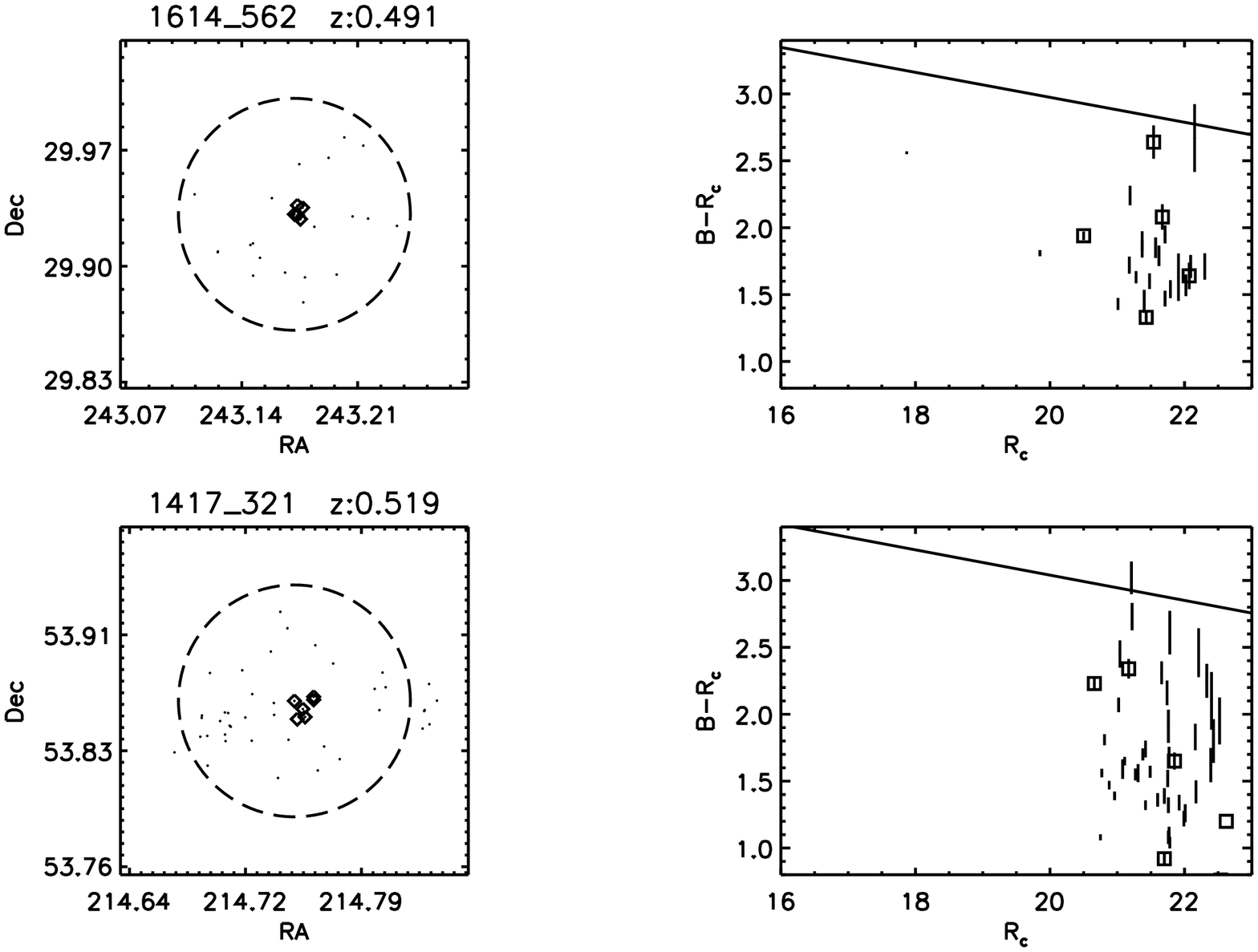}
        \includegraphics[angle=90,width=8.2cm]{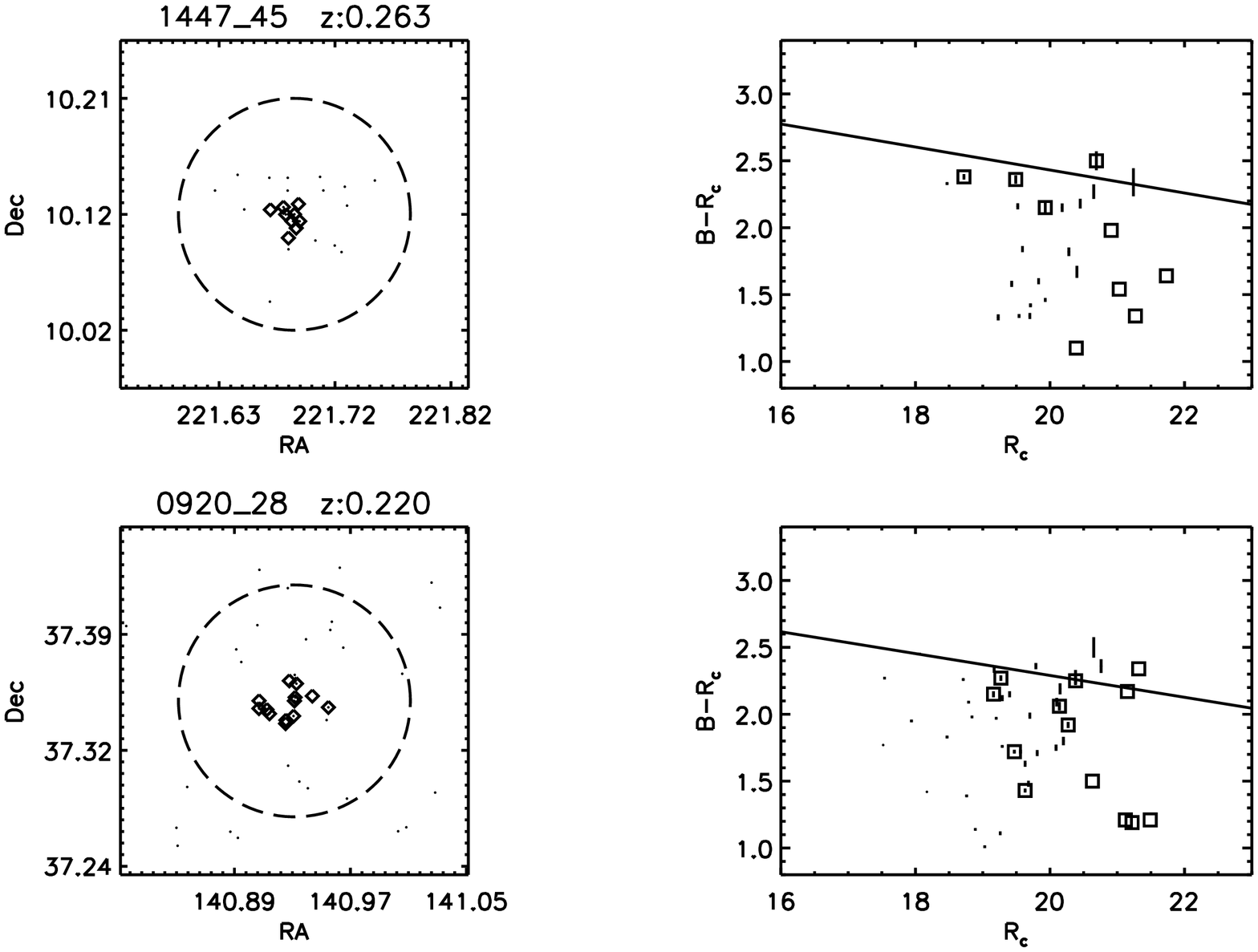}
        \includegraphics[angle=90,width=8.2cm]{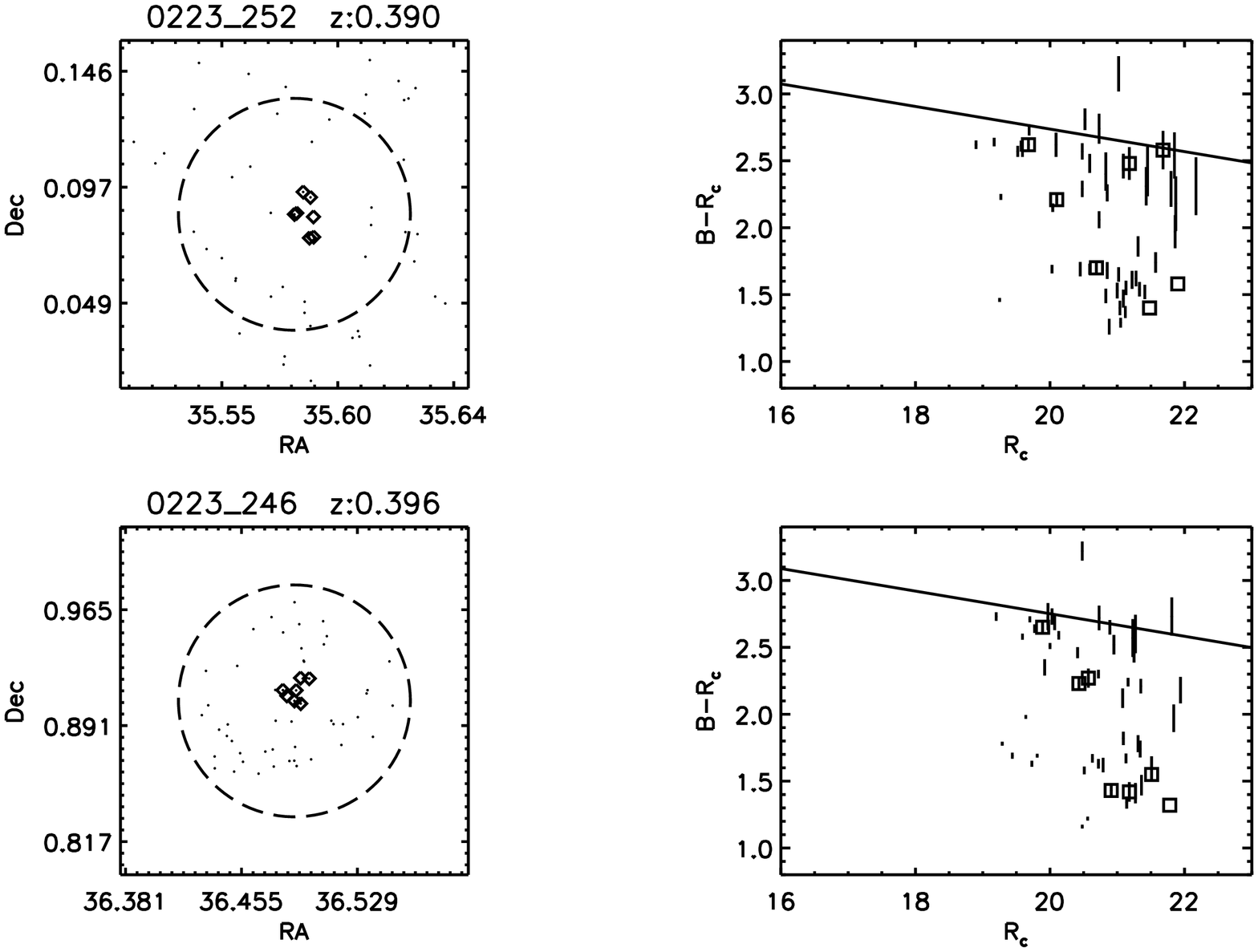}
        \caption{Same as Figure \ref{tmap}. \label{tmap3}}
        \end{figure}

We present the basic data of 24 galaxy groups in Table \ref{grpsample} as examples. They are typical groups in each redshift and $M_{*,grp}$ bins in our analyses in \S\ref{sec:results}.
Figures \ref{tmap}, \ref{tmap2}, and \ref{tmap3} show the sky locations and 
observed color-magnitude diagrams of these groups.
The full sample of 1153 groups is presented in the electronic version of the Table.
Some groups are also found as RCS clusters \citep{2005ApJS..157....1G}. 
We note that most groups, especially the richer ones, exhibit a clear red sequence at the expected theoretical model colors.

\begin{table*}
\begin{center}
\caption{Catalogs of RCS Galaxy Groups$^a$ \label{grpsample}}
\begin{tabular}{llrllcc}
\hline\hline\\
ID    &   R.A.$^b$ & Dec.$^b$ & $z_{grp}$  & $N_{gal}^c$ & M$_{grp}^d$ & $B_{gc}^e$ \\
\hline\\
1447\_48 & 14:47:59.4 & 09:52:46.0 & 0.198$\pm$0.011 & 6.014 & 12.45 & 640$\pm$211\\
2148\_59 & 21:52:33.6 & -05:29:24.9 & 0.277$\pm$0.015 & 7.947 & 12.45 & 271$\pm$174\\
1120\_57 & 11:22:17.1 & 25:58:6.66 & 0.349$\pm$0.019 & 8.082 & 12.41 & 442$\pm$211\\
0920\_317 & 09:22:21.9 & 37:50:15.7 & 0.375$\pm$0.010 & 5.486 & 12.41 & 653$\pm$229\\
1417\_181 & 14:16:54.0 & 52:21:29.4 & 0.487$\pm$0.018 & 6.117 & 12.41 & 674$\pm$250\\
1614\_469 & 16:14:36.1 & 30:25:51.2 & 0.490$\pm$0.014 & 6.512 & 12.41 & 167$\pm$203\\
2316\_15 & 23:14:15.5 & 00:43:42.8 & 0.405$\pm$0.012 & 11.18 & 12.39 & 685$\pm$233\\
2148\_153 & 21:53:54.8 & -05:15:33.7 & 0.443$\pm$0.015 & 6.341 & 12.39 & 454$\pm$223\\
0223\_133 & 02:27:53.7 & 01:14:10.1 & 0.207$\pm$0.010 & 5.961 & 12.02 & 736$\pm$218\\
1417\_23 & 14:14:35.6 & 53:54:13.4 & 0.253$\pm$0.008 & 6.497 & 12.02 & 638$\pm$205\\
0920\_55 & 09:23:0.31 & 37:32:48.4 & 0.380$\pm$0.010 & 8.591 & 12.01 & 808$\pm$243\\
0920\_121 & 09:31:23.5 & 37:50:9.19 & 0.503$\pm$0.018 & 9.267 & 12.01 & 244$\pm$218\\
0920\_365 & 09:29:2.96 & 37:49:43.8 & 0.497$\pm$0.017 & 7.045 & 12.01 & 213$\pm$213\\
1614\_70 & 16:13:30.8 & 30:00:6.82 & 0.365$\pm$0.012 & 8.067 & 12.01 & 443$\pm$205\\
0920\_174 & 09:22:0.56 & 37:54:28.6 & 0.418$\pm$0.011 & 6.094 & 11.96 & 344$\pm$208\\
0920\_322 & 09:28:5.15 & 37:55:16.3 & 0.430$\pm$0.017 & 5.371 & 11.96 & 510$\pm$226\\
1614\_54 & 16:17:18.1 & 30:28:27.9 & 0.400$\pm$0.009 & 7.978 & 11.65 & 366$\pm$203\\
2316\_41 & 23:16:59.4 & -00:12:32.5 & 0.442$\pm$0.016 & 10.36 & 11.64 & 395$\pm$214\\
1614\_562 & 16:12:41.3 & 29:55:42.9 & 0.491$\pm$0.019 & 5.272 & 11.62 & 207$\pm$207\\
1417\_321 & 14:19:0.69 & 53:51:41.1 & 0.519$\pm$0.023 & 6.664 & 11.61 & 259$\pm$223\\
1447\_45 & 14:46:45.5 & 10:07:3.18 & 0.263$\pm$0.010 & 6.237 & 11.58 & 188$\pm$173\\
0920\_28 & 09:23:42.5 & 37:20:13.6 & 0.220$\pm$0.009 & 5.666 & 11.57 & 356$\pm$175\\
0223\_252 & 02:22:19.8 & 00:05:12.9 & 0.390$\pm$0.011 & 5.754 & 11.56 & 334$\pm$205\\
0223\_246 & 02:25:57.5 & 00:54:40.9 & 0.396$\pm$0.016 & 5.206 & 11.56 & 264$\pm$200\\
\hline\hline\\
\end{tabular}
\end{center}
\footnotetext[a]{The photometric catalogs are available in \citet{2005ApJS..158..161H} }.\\
\footnotetext[b]{in J2000.} \\
\footnotetext[c]{$N_{gal}$: net pFoF member count after correcting for background contamination and completeness weights.} \\
\footnotetext[d]{$M_{grp}$=log($M_{*,grp}/M_{\sun}$), computed using all group members within 0.5$R_{200}$ and with background contamination corrected.} \\
\footnotetext[e]{$B_{gc}$: computed using all galaxies within 0.25Mpc to the group centers}.
\end{table*}

\begin{figure}
                \includegraphics[width=8.2cm]{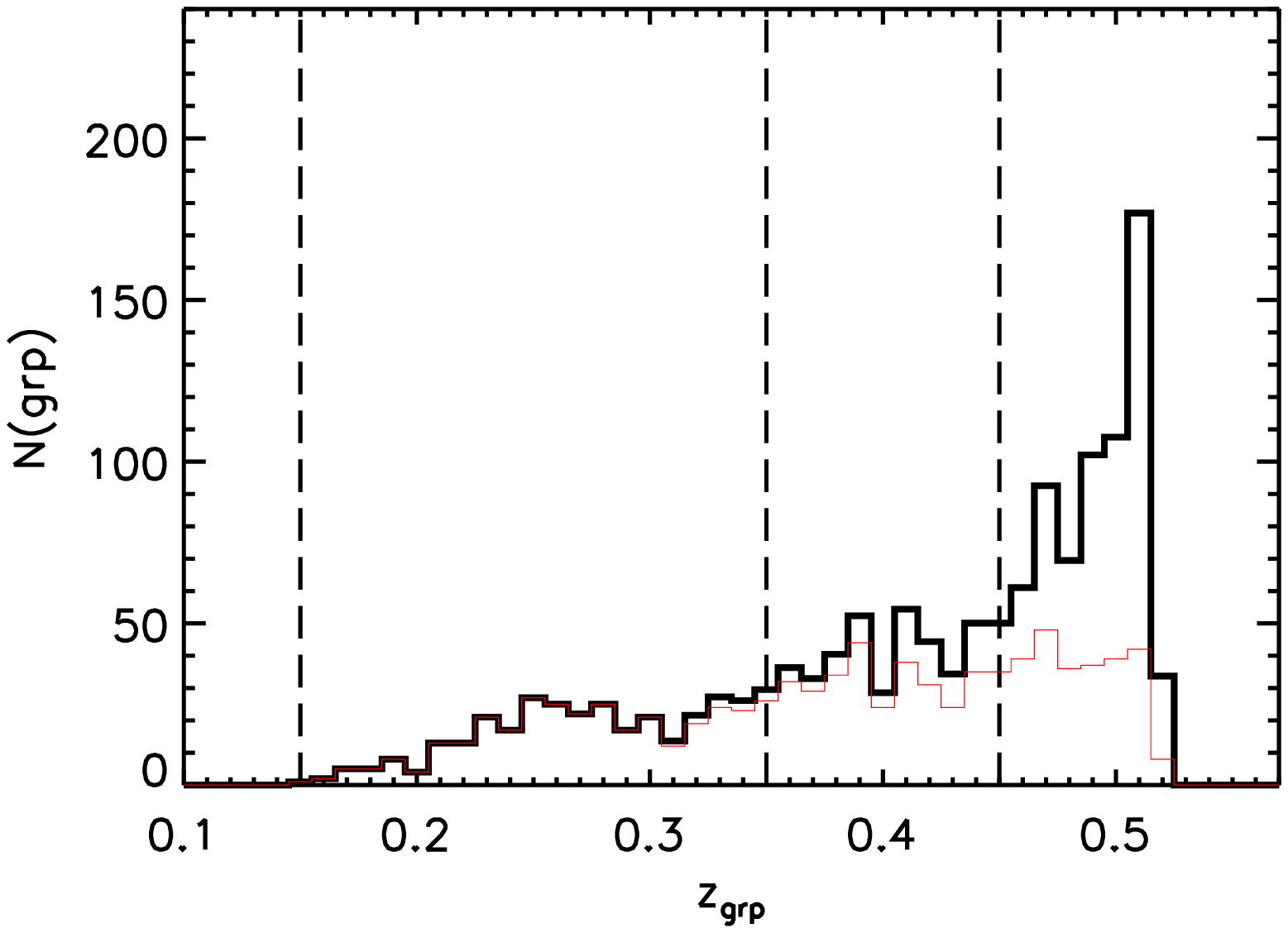}
                \caption{The redshift distribution of 905 galaxy groups in our sample using a 0.01 bin size in $z$.
The thin red and thick black histograms are the distributions before and after applying area corrections.
\label{grpzhistogram}}
                \end{figure}

The redshift distribution of our group sample is plotted in Figure \ref{grpzhistogram}.
Since different patches have different redshift limits, we also
plot in Figure \ref{grpzhistogram}~the redshift distribution corrected
for the sampling areas for the different redshift bins.
For our analysis, we divide the groups into three redshift bins: 
$0.15 \leq z < 0.35$, $0.35 \leq z < 0.45$, and $0.45 \leq z < 0.52$
designated as the $z\sim 0.25$, $z\sim 0.4$, and $z\sim 0.5$ bins in our analysis.
There are 304, 317, and 284 groups, respectively, in these bins.

\subsection{Samples of Group Galaxies}
To select galaxies in the same redshift space as a group, we follow the method used by L09 for selecting galaxies in the same redshift space as the CNOC1 clusters.
That is, the photometric-redshift probability for a galaxy to be at the same
redshift as a given group has to satisfy 
a threshold, which is set to be the same as the redshift linking criterion in the pFoF algorithm.
These galaxies are called `group galaxies' in our analyses 
(\S\ref{sec:results})
and are not limited only to the galaxies linked by the pFoF algorithm, 
since redshift is the  only criterion used for their inclusion.
In essence, the main task of of the pFoF algorithm is to identify
the locations where there are groups; the use of the linkage 
criterion in redshift then generates a sample of galaxies
whose photometric redshifts are consistent with being in these groups.
In our analyses, we further impose group-centric radius criteria for the selection of group galaxies in various radial bins.
We have in total 5028, 5832, and 5560 group galaxies (before background 
subtraction) brighter than $M^*_{R_c}+1.5$ within 0.5$R_{200}$ (derived from group richness,
see \S\ref{rgrp}) in the three redshift bins.

We also estimate the stellar mass for each group galaxy using an empirical 
relation: log(${\rm M}_*/L_{R_c}) = -0.523+0.683(M_{B}-M_{R_c})$, from
 \citet{2003ApJS..149..289B} with solar units from \citet{1994ApJS...95..107W}.
The $M_{B}$ and $M_{R_c}$ are the k-corrected rest-frame absolute
 magnitudes derived using the same method as L09. 
The k-correction values used for each galaxy are based on its
 color and photometric redshift.

\begin{figure}
                \includegraphics[width=8cm,angle=90,scale=0.4]{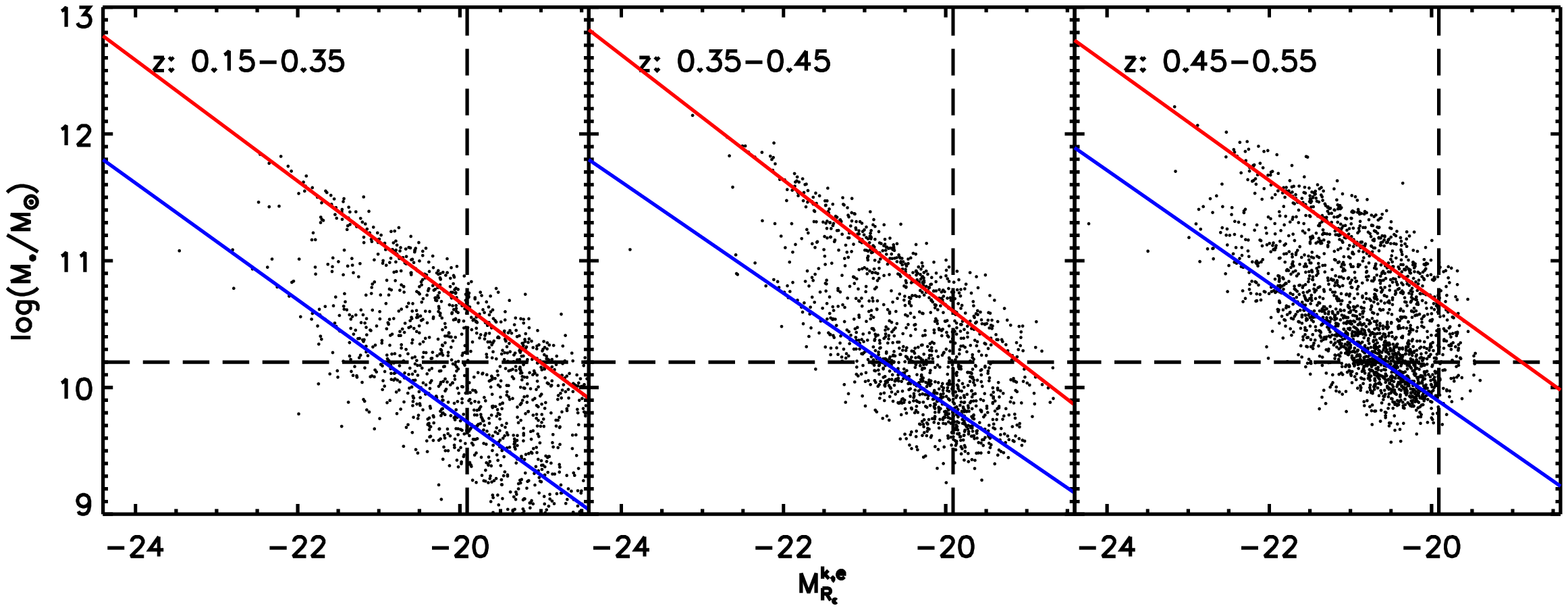}
                \caption{Stellar mass M$_*$ and $M_{R_c}^{k,e}$ for group galaxies within 0.5$R_{200}$ from 182 groups in the RCS $0920$ patch.
The vertical dashed line marks $M_{R_c}^*+1.5$ while the horizontal one, log(M$_*$/M$_{\sun}$)=10.2.
The red and blue solid lines mark the ridge lines for red and blue galaxies, respectively.
Selecting the sample using stellar mass rejects many blue galaxies.
                \label{mag_mass}}
                \end{figure}

Figure \ref{mag_mass} presents the plot of absolute magnitude versus 
stellar mass for our group galaxy sample.
The galaxy sample is complete to $M^*_{R_c}+1.5$~in all three
redshift bins.
In stellar mass, the sample is essentially complete to 
log($\rm M_*/M_{\sun}$)$=$10.2 for the $z\sim0.25$ and $z\sim0.4$ samples;
while the $z\sim0.5$ sample is 100\% 
complete only to log($\rm M_*/M_{\sun}$) $\sim$10.8.
We note that the level of incompleteness at $z\sim0.5$ is corrected substantially by the $w_i$ weight assigned to each galaxy.
There are a total of 3550, 3933, and 3940 group galaxies
within 0.5$R_{200}$ with log($\rm M_*/M_{\sun}$)$\geq$10.2
in the 3 redshift bins, respectively.
The number counts for group galaxies at $0.5\leq R_{200} < 1.0$ are
2637, 3298, and 4605 galaxies for the same redshift divisions.

\subsection{Group-Centric Radius and Scaling Radius}\label{rgrp}
To define the center of a pFoF group, we first take the median R.A. and Dec. of all the linked members as the `fiducial' group center. 
Each linked galaxy is assigned a score based on its $R_c$ magnitude and distance to the fiducial center. 
The score is then computed as the quadratic sum of the ranks of the
magnitude and distance differences between the fiducial
 group center and the galaxy.
Both the magnitude and distance contribute the same weight to the score. 
The galaxy which has the best score is defined as the `central galaxy'.
We adopt the position of the central galaxy as the final center position
of the group from which a group-centric radius is defined.
If more than one linked galaxy in the same group has the same score, we choose the center as the one with the brighter magnitude.

For each group galaxy, we compute the projected distance to the center and scale it by $R_{200}$, the radius within which the density is 200 times the critical density, so that our analysis is not limited only to either the cores of rich groups or the outskirts of poor ones.
The $R_{200}$ of each group is estimated from the correlation between cluster richness $B_{gc}$ (see  \S 3.4) and $R_{200}$ derived using the X-ray luminous CNOC1 clusters in \citet{2003ApJ...585..215Y} (also see \citet{2007ApJ...671.1471B}).
The typical $R_{200}$ in our RCS group sample is $\sim 1.07\pm0.24$Mpc.
The group-centric radius is denoted by $r_{grp}$ in units of $R_{200}$.
We note that the $B_{gc}-R_{200}$ relation for low-mass galaxy groups may not be identical to that of the CNOC1 clusters, but it still gives us an estimate
and rank-order of $R_{200}$.  The effects of the uncertainties in 
the $R_{200}$ on our results are further discussed 
in \S5.1.3.  

\subsection{Group Richness and Masses} \label{Bgc}
Galaxy groups and clusters with more members are likely associated with more massive dark matter halos and have deeper gravitational potentials \citep[e.g.,][]{2003ApJ...585..215Y, 2007A&A...464..451P}.
Therefore, to investigate the influence caused by galaxy groups, the group richness should be considered.
We explore the parameterization of  group richness using two different
parameters: the total stellar mass within a fixed fraction of $R_{200}$ and the richness parameter $B_{gc}~$\citep{1979MNRAS.189..433L}.
 
\begin{figure}
                \includegraphics[width=8cm,angle=90,scale=0.8]{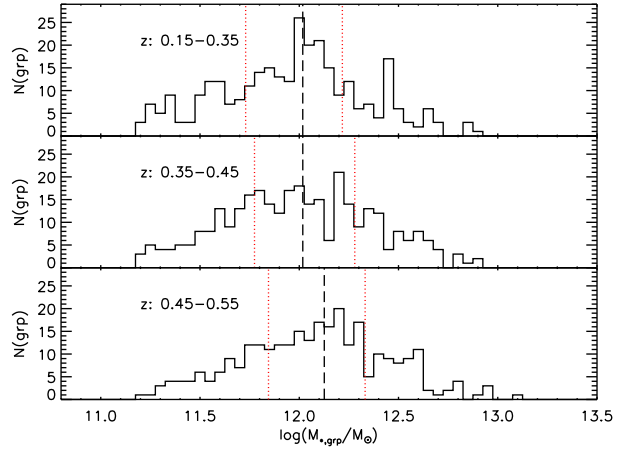}
                \caption{The histogram of total stellar mass within 0.5$R_{200}$ in each galaxy group. The vertical dashed line is the median value, and the two (red) dotted lines are the 25 and 75 percentiles.
The median total group stellar mass remains similar over the three redshift bins.
                \label{mass_histogram}}
                \end{figure}

We compute the total stellar mass of a galaxy group, M$_{*,grp}$, using
 the sum of the stellar masses of all the group galaxies within 0.5 $R_{200}$
in the log (M$_*/M_{sun}$)$\ge$10.2 sample.
The M$_{*,grp}$ is corrected for background contamination (see section \S\ref{sec:background}).
We use 0.5$R_{200}$ instead of 1$R_{200}$ because group member counts have a
larger excess relative to the background within the smaller radius.
We note that group galaxies in the $z\sim0.5$ bin are incomplete 
at log(M$_*/M_{\sun}$)=10.2.
However, partly due to the galaxy weight corrections
and the small contribution to the total group mass from low-mass
galaxies, this has only a few percent effect on the total group
stellar mass. 
This is found to be the case  when the incompleteness effect
 is tested using the average ratios of  M$_{*,grp}$
computed with limits of log(M$_*/M_{\sun}$)=10.2, 10.6, and 11.0 for
each redshift bin.
Figure \ref{mass_histogram} presents the M$_{*,grp}$ distribution of our group sample in each redshift bin. 
The distributions are similar for the three redshift bins, peaking
at or slightly above log(M$_*/M_{\sun}$)=12.0, with median values differing
by $\sim$0.1 dex.
The distributions also show that the sample is beginning to be
incomplete at log(M$_{*,grp}/M_{\sun}$)$\ltapr$12.
This is likely due to the pFoF algorithm's inability to find loose
poor groups, and conclusions based on analyses of the low-mass groups
should take this bias into consideration. 

The $B_{gc}$ parameter is defined as the amplitude of the cluster center-galaxy correlation function, 
and can be estimated via a deprojection of the angular correlation function onto the spatial one assuming spherical symmetry \citep{1979MNRAS.189..433L}.
It is estimated by counting excess galaxies up to a certain absolute magnitude within a given radius, and is corrected for the background counts and scaled by an average luminosity function and average spatial profile. 
Details on its computation, robustness as a cluster richness parameter,
and as a mass indicator are presented
in \citet{1999AJ....117.1985Y} and \citet{2003ApJ...585..215Y}.
 
 	For the RCS-1 clusters,  $B_{gc}$ is computed in a 0.50 Mpc $h_{50}^{-1}$ radius (Gladders \& Yee 2005). 
 Considering that galaxy groups are poorer systems than galaxy clusters, 
 we calculate $B_{gc}$ for each galaxy groups using a 0.25 Mpc $h_{50}^{-1}$ radius to minimize background noise.
 We find that the $B_{gc}$ values computed using a radius of 0.50Mpc are $\sim 35\%$ smaller than those using a 0.25Mpc radius, indicating a steeper slope of the galaxy correlation function than the canonical $\gamma=1.8$. 
 We note that a cluster of $B_{gc}=600$Mpc$^{1.8}h_{50}^{-1.8}$ (within a radius of 0.5Mpc) has a richness equivalent to Abell class 0, or a mass of $\sim3\times 10^{14} h^{-1}
 M_{\sun}$ \citep{2003ApJ...585..215Y}.

\begin{figure}
                \includegraphics[width=8cm, angle=90,scale=0.8]{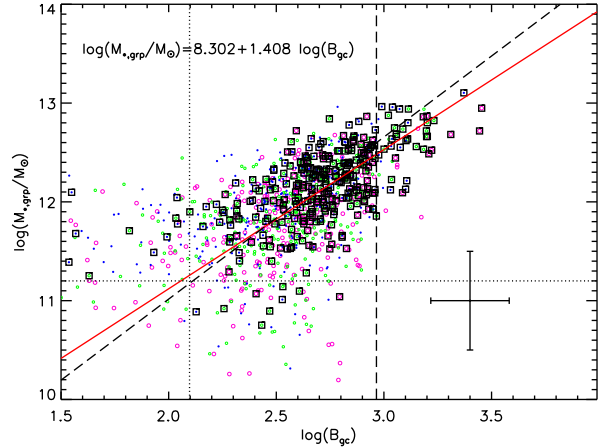}
                \caption{The comparison between M$_{*,grp}$ and $B_{gc}$,
two parameters used as indicators of group richness.
A correlation is seen.
The dotted lines indicate the $B_{gc}$=125 and log($M_{*,grp}/M_{\sun}$)=11.2
criteria for selecting our final group sample.
The solid line is the linear fit (displayed on the top of the figure)
between log($M_{*,grp}/M_{\sun}$) and log($B_{gc}$) using groups with $N_{gal}\geq 8$, while the dashed line is the  fitted relation using a slope of 1.64
(the slope obtained by
Yee \& Ellingson (2003) fitting $M_{200}$ vs $B_{gc}$ for a sample of
CNOC1 clusters).
The vertical dashed line marks the mass of $M_{200}\sim3\times10^{14}$ \msun,
based on the calibration of Yee \& Ellingson.
Galaxy groups are represented by circles, with larger symbols representing
groups at lower redshift.
Squares mark groups with linked pFoF members $N_{gal}\geq 8$.
The error-bars on the lower right represent typical uncertainties in the two
parameters.
\label{bgc_mass}}
                \end{figure}

Figure \ref{bgc_mass} presents the comparison between $B_{gc}$ and M$_{*,grp}$.
A positive correlation is observed, with the scatter broadly consistent
with the measurement errors.
We obtain the best fitting power-law, using the more robust 
$N_{gal}\ge8$ sample:
${\rm log}( {\rm M}_{*,grp}/M_{\sun}) =1.41 {\rm log } (B_{gc})+8.30$.
Using the simple scaling relations in \citet{2003ApJ...585..215Y} with the assumption that galaxies trace total halo mass,
we should expect the halo mass within a dynamically scaled radius
(e.g., $M_{200}$, the mass within $R_{200}$) to scale as $B_{gc}^{3/\gamma}$,
where $\gamma$ is the power-law slope of the galaxy-galaxy 
correlation function; or, for the canonical $\gamma$ of 1.8, ${\rm log} 
(M_{200})\sim B_{gc}^{1.66}$.
Thus, our power-law fit of ${\rm M}_{*,grp}$ and $B_{gc}$
is consistent with ${\rm M}_{*,grp}$ being a direct one-to-one proxy for the
$M_{200}$ halo mass of the groups.
Using the correlation from Yee \& Ellingson (2003) between $B_{gc}$
and $M_{200}$, and compensating for $H_0$, we can roughly estimate
the relation between the total group stellar mass and the halo mass.
We do this by fitting the M$_{*,grp}$--$B_{gc}$ relation using the
same power-law slope (1.64) as that measured by Yee \& Ellingson (2003)
for the $M_{200}$--$B_{gc}$ relation for the CNOC1 clusters,
and correcting for factor of 0.65 in
$B_{gc}$ values computed using a 0.5 $h_{50}^{-1}$ Mpc radius.
This gives 
log($M_{200,grp}$)$\sim$log($M_{*,grp}$)+1.85.
This provides a very rough scaling, probably no better than
a factor of two, between the group stellar mass and
 $M_{200}$.
In this scaling, a group with 
log($M_{*,grp}/M_{\sun}$)=12.0,
(approximately the median \SMgrp~for our sample),
represents a halo mass of $\sim 7 \times 10^{13}\msun$.
In the remainder of the paper, we use the group stellar mass M$_{*,grp}$ as the richness indicator and a rough halo mass proxy. 

\subsection{Background Correction} \label{sec:background}
Because of the relatively large uncertainty of 
the photometric redshift, even with the use of the pFoF algorithm for
determining whether a galaxy belongs to a given group, there  will
still be a significant fraction of ``group galaxies" which are
field, or near-field, galaxies, especially at the larger group-centric radii.
Hence, it is essential to perform background contamination corrections
in analyses using the group galaxy sample.
We construct both luminosity-limited and stellar-mass-selected samples
of background counts using three RCS patches: \it 0920 \rm, \it 1417 \rm,and \it 1614 \rm. 
These three patches have the deepest photometry 
and cover a total of 11.38 deg$^2$ on the sky. 
We note that these three control patches are part of our whole sample.
The deep photometry allows us to compute galaxy counts to a higher redshift.

We use the method in L09 to construct the galaxy number surface density 
per Mpc$^2$ in the field in redshift bins of $\Delta z=0.01$, 
denoted as $n(z,M_{R_c})$ and $n(z,$M$_*)$. 
Briefly, the $n(z)$ functions are obtained by integrating the sum of
weighted photometric redshift probability densities of all galaxies
in the three patches within each
small redshift interval and normalized by the area of the patches.
For a given set of galaxies, e.g., a sample of group galaxies within
certain group-centric radius over certain magnitude range,
the total number of background galaxies
is computed by multiplying the appropriate $n(z)$ curve with the summed 
photometric-redshift likelihood function of all galaxies of interest,
normalized by the area.
A more detailed description of the method can be found in L09.

\subsection{Local Galaxy Density}
It has been shown that local galaxy density plays a role in transforming star-forming galaxies into passive ones \citep[e.g.,][]{2003ApJ...584..210G,2003PASJ...55..757G}.
We compute local galaxy density $\Sigma_5$ based on the nearest 5$^{th}$ neighbor galaxy for each group galaxy using the `group galaxies' in our two samples. 
The $\Sigma_5$ is then corrected for background galaxy contamination (see L09 for more details of the computation).
The $\Sigma_5$ is computed using galaxies brighter than $M_{R_c}^* + 1.5$ for the luminosity-limited sample, and using galaxies with log$(M_*/M_{\sun}) \geq 10.2$ for the stellar-mass-selected sample.
We note that in general blue galaxies in the stellar-mass-selected sample have
somewhat smaller $\Sigma_5$ than those in the luminosity-limited sample, and the differences are primarily seen in low $\Sigma_5$ regions.  
This is because most blue galaxies in the luminosity-limited sample have stellar mass less than log$(M_*/M_{\sun})=10.2$, and they are excluded in the stellar-mass-selected sample.

\subsection{Red Galaxy Fraction} \label{fred}
As the red sequence of early-type galaxies is the final destination for galaxies on a color-magnitude diagram, 
we probe the dependence of the red galaxy fraction, $f_{red}$, with various group properties.
The $f_{red}$ is computed based on the same method as L09 but using samples of  galaxies defined by  stellar-mass limits.
The red galaxies are defined as those whose colors are redder than half the color differences between the spectral energy distributions (SEDs) of E/S0 and Sc galaxies at a given redshift,
which is tabulated in \citet{1997A&AS..122..399P}.
This color division changes with redshift on an observed color-magnitude 
diagram, and follows the passive evolution of galaxy colors.
The color division is typically 0.25 to 0.30 mag to the blue of the red
sequence, and is comparable to the  $\Delta B-V=0.2$ used by Butcher \&
Oemler (1984), and the definition used by \citet{2001ApJ...547..609E}.
The background count corrections for the red galaxy counts are derived using
the method described in \S3.5 but applied to background galaxies with the 
color cut described above.
We compute $f_{red}$ using a statistical inference \citep{2004physics..12069D,2006MNRAS.365..915A}
because the background contamination plays a role in estimating the true fraction of red group members especially when the number of galaxies in a group is not large.

\section{Results} \label{sec:results}
In the following subsections, we investigate $f_{red}$ as a function of redshift, M$_{*,grp}$, $r_{grp}$, and $\Sigma_5$.
We have explored results using both the luminosity and stellar-mass samples. 
We find that both samples exhibit consistent trends and lead to similar conclusions. 
We therefore present and discuss our results using only the stellar-mass selected sample. 
We note that, as discussed in L09, using a luminosity-selected galaxy sample is equivalent to having different stellar-mass limits for the red and blue galaxies.
Our luminosity-selected sample has equivalent stellar-mass limits of log(M$_*$/$M_{\sun}$)$\sim10.5$ and log(M$_*$/$M_{\sun}$)$\sim9.8$, for red and blue galaxies (Figure \ref{mag_mass}), respectively.
On the other hand, the nominal stellar-mass limit of log(M$_*$/$M_{\sun}$)=10.2
of the stellar-mass selected sample 
corresponds to $M_{R_c}^{k,e}\sim-19.0$ and --20.75, for red and blue
 galaxies, respectively.
Thus, for the $z\sim0.5$ bin the red group galaxy sample is substantially
incomplete at log(M$_*$/$M_{\sun}$)=10.2, and we include results from the 
$z\sim0.5$ bin only for samples with a stellar-mass
limit of log(M$_*$/$M_{\sun}$)$\ge$10.6.

To investigate the dependence of galaxy population on group richness or mass,
we also divide our group sample into high-, intermediate- and low-mass
bins of log(M$_{*,grp}/M_{\sun}{\rm )}\ge12.2$, 12.2$>$log(M$_{*,grp}/M_{\sun}$)$\ge$11.8, and 11.8$>$log(M$_{*,grp}/M_{\sun}$)$\ge$11.2.
These bins correspond to, approximately, 
$M_{200}\ge1.1\times10^{14}M_{\sun}$,
$1.1\times10^{14}M_{\sun}> M_{200} \ge4.5\times10^{13}M_{\sun}$,
and $4.5\times10^{13}M_{\sun}> M_{200} \ge 0.7\times10^{13}M_{\sun}$,
respectively.
We intend to probe $f_{red}$ out to $r_{grp}\sim2$. 
Due to the less accurate redshift information in photometric redshifts, a small 
number of galaxies may belong to more than one group. 
In this situation, we re-assign the galaxy's group membership to the group for which the group-centric radius is the smallest.

\subsection{Color-magnitude Diagrams and the Red-Sequence for Group Galaxies}
\begin{figure}
                \includegraphics[width=8cm,angle=90,scale=0.8]{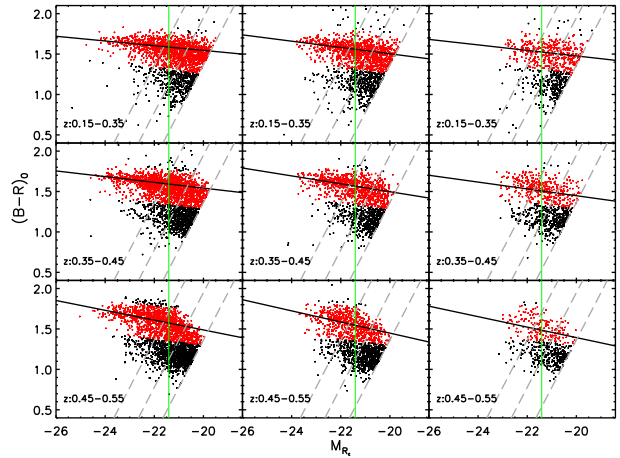}
                \caption{
Stacked rest-frame $B-R$ vs $M_{R_c}$ for group galaxies within 0.5$R_{200}$ in each redshift and $M_{*,grp}$ bin.
The gray dashed lines in each panel mark log($M_*/M_{\sun}$)=11.0, 10.6, and 10.2 from left to right.
The vertical green line is $M_{R_c}$=21.41.
Red dots represent galaxies satisfying the definition of red galaxies as
described in \S3.7.
The solid lines are the fitted red sequences using red galaxies with log($M_*/M_{\sun}$)$\geq$10.6.
The fitting results are listed in Table \ref{tab:redsequence}.
Group galaxies, including those in poor groups, exhibit similar color-magnitude relations as those in clusters.
                \label{cmr}}
                \end{figure}

Figure \ref{cmr} presents the stacked  rest-frame  color-magnitude diagrams for group galaxies within 0.5$R_{200}$ at each redshift and M$_{*,grp}$ bin.
We do not apply any background correction in Figure \ref{cmr}, 
but the contamination of background galaxies is not expected to be large
for reasonably luminous galaxies,
since these group galaxies are already pre-selected by photometric redshift
and a relatively small group-centric radius is used.
The very small number of galaxies in the CMDs redder than the
red-sequence indicates that the contamination rate is small. 
We find that group galaxies exhibit color-magnitude distributions
similar to clusters:
that is, group galaxies can be separated into  two distinct
populations of red and blue galaxies, and the red members,
including those in poor groups, form a clear sequence. 

We fit the red sequence using red galaxies with log($M_*/M_{\sun}$)$\geq$10.6, and present the resulting fit in Table \ref{tab:redsequence}.
\begin{table} 
\caption{The Fitted CMR for Fig.\ref{cmr} \\
in the form as $y$ = A + B$x$ or $y$ = C + B($x$+21.41) 
} \label{tab:redsequence}
\begin{tabular}{lllll}
\hline\hline\\
\hline\\
$z$        & $M_{*,grp}$ & A & B & C \\
0.00--0.35 & 12.2--14.0 & 0.963$\pm$0.049 & -0.029$\pm$0.002 & 1.584 \\
0.00--0.35 & 11.8--12.2 & 0.723$\pm$0.068 & -0.039$\pm$0.003 & 1.558 \\
           & 11.2--11.8 & 0.796$\pm$0.001 & -0.034$^{*}$ & 1.524 \\
0.35--0.45 & 12.2--14.0 & 0.834$\pm$0.039 & -0.035$\pm$0.002 & 1.583 \\
0.35--0.45 & 11.8--12.2 & 0.520$\pm$0.098 & -0.049$\pm$0.004 & 1.569 \\
           & 11.2--11.8 & 0.605$\pm$0.002 & -0.042$^{*}$ & 1.504\\
0.45--0.55 & 12.2--14.0 & 0.274$\pm$0.075 & -0.061$\pm$0.003 & 1.580 \\
0.45--0.55 & 11.8--12.2 & 0.071$\pm$0.162 & -0.069$\pm$0.007 & 1.548 \\
           & 11.2--11.8 & 0.098$\pm$0.005 & -0.065$^{*}$ & 1.490\\
\hline\hline\\
\end{tabular}
\tablenotetext{*}{using the mean slope of the other two $M_{*,grp}$ bins at the same redshift.}\\
\end{table}

We use a relatively high stellar-mass limit to perform the fit to minimize possible 
confusion arising from the blue cloud galaxies due to larger photometric
uncertainties at the faint end.
These red group galaxies are defined in the same manner as described
in $\S$\ref{fred}.
We further impose a red-cut of 0.3 mag redder than the model red-sequence 
color, in order to minimize the effect on fitting from the small number of 
galaxies in the group galaxy sample which are 
are likely background galaxies with colors significantly redder than
the red sequence model.
Because of the more sparse data for the low stellar-mass groups,
we fix the fitting slope for these groups 
to be that of the average of the two bins of the more massive 
groups at the same redshift to derive the red-sequence colors and dispersion.
We find that the slopes and zero points of the fitted red sequences 
are similar among different redshift and M$_{*,grp}$ bins --
 an indication that the formation epochs for $\sim M^*$ galaxies
in these groups are typically at $z\gtapr2$ 
\citep[e.g.,][]{1998ApJ...492..461S, 1998ApJ...501..571G}.
The average slope of $\sim-0.046$ for the 6 more massive group
samples is consistent with the value of 
--0.047 in the model of \citet{1997A&A...320...41K}.
We see a consistent trend of increasing slope, 
by about --0.03 from the $z=0.25$ to $z=0.5$ bins, which is
entirely in agreement with that observed for
clusters over this redshift range (but for slightly different filters)
in Gladders et al. (1998, see their Figure 4) and their models.

The average red-sequence rest-frame zero point $(B-R_c)_0$ at $M^{*}_{Rc}=-21.41$
is 1.570 mag, and has a small root-mean-squared (rms) dispersion of 0.016
mag over all redshifts and M$_{*,grp}$ bins. 
However, it appears that
there is a trend of a slightly bluer $(B-R_c)_0$ zero point for less
massive groups: The zero points averaged over redshift are  
1.571, 1.576, and 1.564 for the massive-, intermediate-, and low-mass
group samples.  These differences account for almost all the dispersion
in the red-sequence color zero points, as the variations over redshift
within each group mass grouping are only $\sim\pm0.01$ mag.
It is not clear whether this small observed difference is a reflection of a trend in the average age since formation or the average metallicity in the luminous red-sequence galaxies in groups of different masses.

We also compute the dispersion of the rest-frame $B-R_c$ colors 
for the bright ($\leq M^*$) red-sequence galaxies in the 9 subsamples. 
They range from 0.08 to 0.10
at the low- and high-redshift bins for the more massive groups, to about
0.10 for the low-mass groups at all redshift bins.
However, there are two observational uncertainties which contribute
to the dispersion of the red-sequence.
First, the photometric redshifts of the groups have a typical
uncertainty of 0.03, which translates into an uncertainty of about 0.03
mag in the k-corrections for red galaxies.  
Second, the typical mean photometric uncertainties in $B-R_c$ range from 0.03 
mag for  the low-redshift samples (at the brighter end of our photometry) 
to 0.06 mag for the high-redshift samples.
Rough corrections in quadrature for the dispersion due to these effects
give the intrinsic red-sequence color dispersion to be $\sim0.03$ for
all cases.
While the rough corrections and the possible contamination
from background galaxies prevent us from discerning any pattern
in the red-sequence dispersions for the different subsamples, the
estimated average dispersion is consistent with observations of much smaller
samples of rich clusters with more precise redshift and photometric information
\citep[e.g.,][]{1978ApJ...225..742S,1992MNRAS.254..601B, 1997ApJ...483..582E}

This high degree of uniformity in the red sequence in our group samples
suggests that the more
massive ($\sim M^*_{Rc}$, or log(\smass)$\geq$ 11.0) red-sequence galaxies 
are very similar by $z\sim0.5$ in all
groups with a halo mass larger than approximately 10$^{13.5}M_{\sun}$;
i.e., their formation time is likely to be at $z\geq2$,
similar to the conclusions drawn from more massive clusters
\citep[e.g.,][]{1998ApJ...501..571G,2007ApJ...654..138Q}.
 From a visual inspection of the CMDs,
the number of blue group galaxies appear to drop significantly
from $z\sim0.50$ to $z\sim0.25$.
In the next section, we quantitatively measure the fraction of red group 
member galaxies (\fred) and examine their dependence on group and
galaxy properties.

\subsection{Dependence of $f_{red}$ on Group Total Stellar Mass \label{grp-downsize}}
\begin{figure}
                \includegraphics[width=8cm,angle=90,scale=0.45]{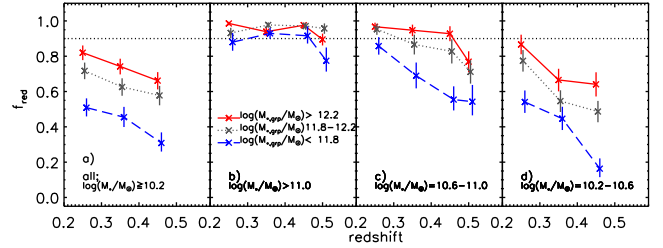}
                \caption{ Red-galaxy fraction, $f_{red}$, as a function of redshift in three log(M$_{*,grp}$) bins using all groups galaxies within 0.5$R_{200}$ without and with stellar mass controlled.
 Groups with larger M$_{*,grp}$ exhibit larger $f_{red}$,indicating an apparent group down-sizing effect.
                \label{fmstar.z}}
                \end{figure}

We probe the red galaxy fraction $f_{red}$ in galaxy groups, computed within 0.5$R_{200}$, as a function of redshift for the three different M$_{*,grp}$ samples and present the results in Figure \ref{fmstar.z}.
Panel {\bf a}) of Figure \ref{fmstar.z} shows  the change in \fred~as a 
function of redshift to $z=0.45$ using a galaxy stellar-mass limit of
log(\smass)=10.2.
Note that we measure \fred~only out to $z\sim0.45$, as at $z\sim0.5$
the galaxy sample is not complete at the log (\smass)=10.2 level.
All three M$_{*,grp}$ subsamples exhibit increasing $f_{red}$ toward lower redshift; i.e., the Butcher-Oemler effect \citep{1984ApJ...285..426B} observed on the scale of galaxy groups.
The more massive groups have a larger $f_{red}$ in all redshift bins.
In other words, more massive groups turn their galaxies red, due to the suppression of star formation activity, at an earlier time than those in less massive groups. 
This result can be described as a down-sizing evolution 
effect of sorts operating in galaxy groups.

To ascertain that this effect is not the result of relatively more massive galaxies inhabiting more massive groups, 
we further control for the stellar mass of group member galaxies in the next
three panels of Figure~\ref{fmstar.z}.
To aid in the discussion of the evolution of \fred, we use 
\fredT=0.85 as the threshold \fred~for which a sample of galaxies 
is considered as
having the bulk of its galaxies with their star formation quenched.
Panel {\bf b}) shows that the most massive group member galaxies 
(log($M_*$/$M_{\sun}$)$>$11) exhibit essentially flat \it $f_{red}$-z \rm 
trends for groups in different M$_{*,grp}$ bins with \fred$>$\fredT, 
with the 
exception of the $z\sim0.5$ point for the low-mass groups. 
That is, these massive group galaxies have mostly finished their 
star formation by $z\sim0.5$, regardless of the influence of group environment.

In contrast, the effect of M$_{*,grp}$ is clearly observed for the subsamples 
of galaxies of intermediate and low stellar masses. 
First, galaxies of similar stellar masses are redder in more massive groups
(panels {\bf c} and {\bf d}), regardless of redshift, demonstrating 
the dependence on \SMgrp.
Second, the \it $f_{red}$-z \rm 
trends are different for intermediate- and low-mass galaxies in groups 
of different \SMgrp.
The intermediate-stellar-mass galaxies in the massive groups reach
\fredT~at $z$ between 0.45 to 0.25, depending on the mass of the
groups to which the galaxies belong.
For the low-stellar-mass galaxies, only those in the most massive
groups manage to barely reach \fredT~at $z\sim0.25$.
In general, the low-mass galaxies also show larger changes in \fred~than the
intermediate-stellar-mass galaxies over $z\sim0.45$ to $z\sim0.25$ range.
The different behaviors of the \it $f_{red}$-z \rm trends for these three 
subsamples indicate that, while the \fred~correlates strongly with the
 stellar mass of the group galaxies, the halo mass of the group in which
the galaxies are situated also has a significant influence on the evolution
of the lower-mass group galaxies. Galaxies in higher mass groups 
reach \fredT~at an earlier epoch and have a less steep $f_{red}$-$z$ trend.

The large \fred~and flat trends with redshift for the massive group galaxies
further provides evidence of  galaxy `down-sizing' studied in the literature;
i.e.,  star formation takes place and also stops earlier in massive galaxies, then shifts to less massive systems at later times \citep[e.g.,][]{1996AJ....112..839C,2008MNRAS.389..567C,2008MNRAS.386.1695S}.
Massive galaxies, by and large, show little effect from being
in groups of different masses (at these moderate redshifts); 
their evolutionary stage is likely a
reflection primarily of their own mass.
However, lower-mass galaxies do show evolutionary histories significantly
correlated with the halo mass of their parent groups.
Hence, the `group down-sizing' observed in panel {\bf a)} of  Figure
\ref{fmstar.z} is not caused by differences in the stellar mass distributions
 of member galaxies in groups with different richness, but rather is the result
of lower-mass member galaxies in higher-mass groups being in a more
advanced evolutionary state than those in lower-mass groups.

\subsection{Dependence of $f_{red}$ on Group-Centric Radius}        
In the literature, many studies have suggested that group environment can truncate star formation in their member galaxies \citep[e.g.,][L09]{2008MNRAS.tmp.1000F,2008A&A...486....9V, 2008MNRAS.387...79V, 2008ApJ...680.1009W}.
For example, \citet{2002MNRAS.335..825D} have shown, using groups from
the 2dF,  that the fraction of galaxy populations in groups, especially the 
rich and massive ones, exhibit similar radial trends as those observed in galaxy clusters with an offset in the values of \fred.
Using the CNOC2 group catalog, \citet{2005MNRAS.358...71W} found that the
fraction of passive spiral galaxies, which are galaxies of spiral 
morphology but with no/little [OII] emission lines, increases
 continuously toward the group center, and demonstrate that star formation rate is truncated gradually as galaxies move toward the group center.
 
\begin{figure}
                \includegraphics[width=8cm,angle=90,scale=0.4]{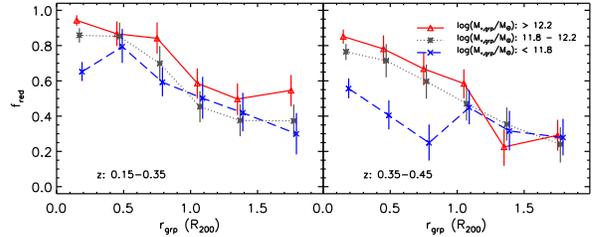}
                \caption{Red-galaxy fraction $f_{red}$ as a function of group-centric radius $r_{grp}$ in units of $R_{200}$ by stacking all group galaxies in each bin. Groups are separated into three richness bins indicated in the plot. The $f_{red}$ decreases with $r_{grp}$. The differences in $f_{red}$ among groups of different richness are seen primarily within the virialized region ($r_{grp} < 1$).
                \label{fgcplot}}
                \end{figure}

To quantify radial color trends within galaxy groups, 
we compute $f_{red}$ as a function of $r_{grp}$.
We separate the group sample by group total stellar mass M$_{*,grp}$, 
since we have found in \S4.2 that the $f_{red}$ within groups at a fixed redshift depends on M$_{*,grp}$.
The results, using group galaxy samples with a stellar mass limit of
log($M_*/M_{\sun}$)$\ge10.2$ are presented in Figure \ref{fgcplot} for the two lower redshift bins.

\begin{figure}
                \includegraphics[width=8cm,angle=90,scale=0.8]{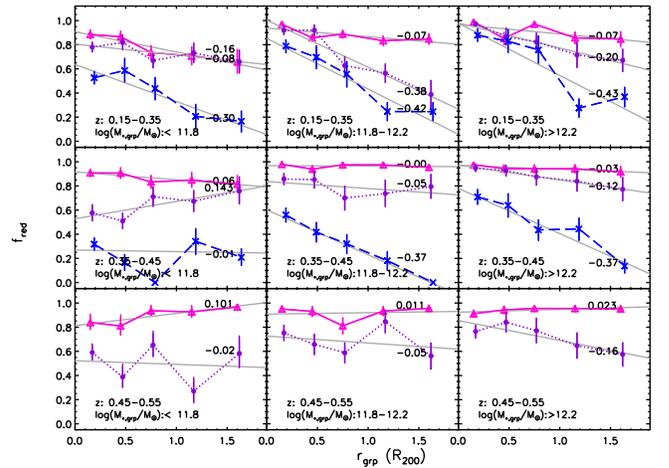}
                \caption{\it `$f_{red}$-$r_{grp}$' \rm trends for group galaxies with log(M$_*/M_{\sun}$) $>$ 11.0 (solid pink), log(M$_*/M_{\sun}$)=10.6-11.0 (dotted purple), and log(M$_*/M_{\sun}$)=10.2-10.6 (dashed blue) in each redshift and log(M$_{*,grp}$) bin as indicated in the panels.
A simple linear fit is applied to each trend with the slope indicated.
In general more massive group member galaxies exhibit gentler slopes toward to the center, and more massive groups produce steeper radial dependence.
                \label{mgcplot}}
                \end{figure}

Generally speaking, group galaxies in all  M$_{*,grp}$ and 
redshift bins exhibit declining $f_{red}$ with increasing $r_{grp}$. 
The central regions of the groups are dominated by red members with
\fred$\sim0.6$ to 0.9, dropping to $\sim$0.3--0.4 at the outskirts. 
For both redshift bins, the effect of M$_{*,grp}$ is clearly observed within 
$r_{grp}\ltapr1$, where groups of higher M$_{*,grp}$ have larger $f_{red}$.
Beyond $R_{200}$, the differences in $f_{red}$ among different M$_{*,grp}$ groups become smaller, and approaching similar values of
\fred$\sim0.3$ to 0.4 for all groups.
However, the observed  trends in Figure \ref{fgcplot} are
likely driven differently by group galaxies of different stellar masses.
We accordingly re-plot the \it $f_{red}$-$r_{grp}$ \rm trends by 
dividing our group galaxies into three subsamples, with log(M$_*/M_{\sun}) > 11.0$, log(M$_*/M_{\sun}) = 10.6-11.0$, and log(M$_*/M_{\sun}) =10.2-10.6$.
For the two subsamples of more massive galaxies, we also include the
$z\sim0.5$ redshift bins.
The results are presented in Figure \ref{mgcplot}.
As an aid to interpretation,
we also perform linear fits of the $f_{red}$-$r_{grp}$ trends
for the 24 subsamples, with the best fitting slopes shown in 
the Figure.

While the error bars are large for these plots, several trends are apparent.
As expected, the most massive galaxies (log(M$_*/M_{\sun}$)$>$11) show 
little effect from their location within their parent group, 
as almost all have essentially \fred$\geq$\fredT.
The changes in the $f_{red}$-$r_{grp}$ trends with $z$ or M$_{*,grp}$ 
are primarily driven by group galaxies with intermediate 
 (10.6$\leq$ log(\smass)$<$11.0), and low masses (10.2$\leq$log(\smass)
$<$10.6).
These $f_{red}$-$r_{grp}$ trends follow a couple of general patterns.
First,  the low-mass
galaxy samples have the steepest radial changes of \fred~for all group
mass and redshift bins, indicating that they are affected by the group
environment the most.
Second, the radial slope of \fred~appears somewhat shallower for these 
galaxies in low-massive groups, which is suggestive that more massive 
halos may have a relatively larger effect on the galaxies that
they are accreting, although larger samples are needed to verify this effect.

\subsection{Local Galaxy Density for Group Galaxies}
It has generally been accepted that local galaxy density can affect galaxy properties such as star formation rates and colors (see \S\ref{introduction}).
Using a small sample of rich clusters,
L09 probed the local galaxy density effect on cluster galaxies at different cluster-centric radii over $0.15 < z < 0.55$. 
They showed that local galaxy density has an important effect at cluster outskirts but less so in the cluster central regions.
Here, with a much larger sample covering a larger range of halo masses,
we examine the interplay between  the
local galaxy density effect and group environmental influence,
as parameterized by the group-centric radius, on group galaxies.
These two measures can be considered different in that the former
tells us how close neighbors affect the galaxy population, while the
latter delineates the effect due to the position of the galaxy in
the larger parent halo.

\begin{figure}
                \includegraphics[width=8cm,angle=90,scale=0.8]{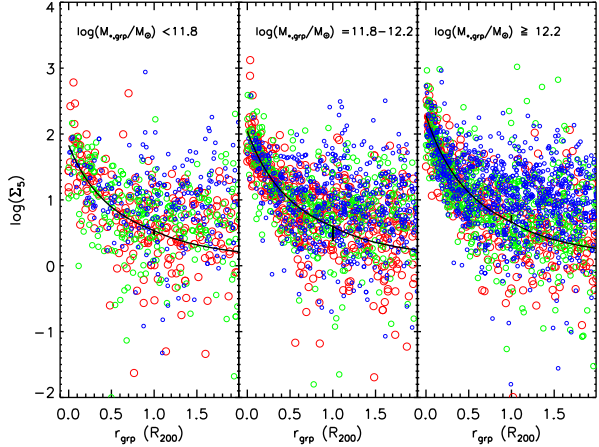}
                \caption{Local galaxy density $\Sigma_5$ as a function of $r_{grp}$ in groups with different M$_{*,grp}$.
Larger circle sizes indicate galaxies at lower redshift.
The solid curves are the fitted 2D NFW \citep{1996ApJ...462..563N} relations using the analytic form in \citet{1996A&A...313..697B}.
                \label{2dnfw}}
                \end{figure}

Figure \ref{2dnfw} plots the local galaxy density, $\Sigma_5$, for
each group galaxy,  which is measured based on the nearest fifth neighbor distance using the method in L09, 
as a function of the group-centric radius for groups with different M$_{*,grp}$.
The distributions cover over three orders of magnitude in local
density and can be described in general as having a decreasing 
average with increasing radius. 
In each M$_{*,grp}$ bin, we also show that the local density distribution
roughly follows a projected NFW profile \citep{1996ApJ...462..563N}, fitted
following the prescription of \citet{1996A&A...313..697B}.
These plots show that the $\Sigma_5$ measurement produces
meaningful relatively local density information when the density
range is sufficiently large, e.g., over a factor of several 
from low to high density.
Typically, the group core has local galaxy densities that are close to 100 to 
several hundred times higher than those at 2$R_{200}$, 
for low- and high-mass groups, respectively.
Within 0.5$R_{200}$, $\Sigma_5$ decreases smoothly with $r_{grp}$.
Beyond that, some regions with higher $\Sigma_5$ are super-imposed on the 
envelope, indicating possible infalling sub-groups at large $r_{grp}$, or
the ambiguity in group membership among multiple groups.

\subsubsection {The Effect of Local Density}

\begin{figure}
                \includegraphics[angle=90,width=8.2cm]{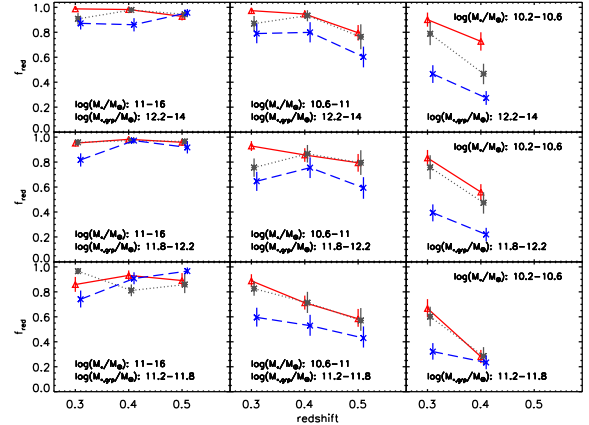}
                \caption{Red-galaxy fraction $f_{red}$ within 1.5$R_{200}$ as a function of redshift for galaxies in different local galaxy density $\Sigma_5$ subsamples.
The 9 panels separate subsamples of different group stellar masses M$_{*,grp}$ and galaxy stellar masses M$_*$, as indicated on each panel.
The red solid, gray dotted, and blue dashed curves represent high-, intermediate-, and low-$\Sigma_5$ divisions, respectively.
                \label{dgcplot.r.3_1.5R200}}
                \end{figure}

To examine the effect of local galaxy density environment on the galaxy
population, in Figure \ref{dgcplot.r.3_1.5R200} we plot \fred~as a function of redshift for group galaxies within 1.5$R_{200}$ into three bins of local galaxy density.
We further separate the galaxy samples into bins of galaxy
stellar mass and total group stellar mass as in \S4.3, 
since we have found that \fred~has a dependence on both of these parameters.
Figure \ref{dgcplot.r.3_1.5R200} provides a look at the effect of 
$\SigmaF$ integrated over the whole group.
Our highest and lowest $\Sigma_5$ bins contain the highest 20-percentile
and the lowest 40 percentile of group galaxies in our sample.
The median local galaxy density values of the high- and intermediate-density 
bins are  approximately 12.8 and 3.7 times that of the low-density bin, 
respectively.

The main feature of Figure \ref{dgcplot.r.3_1.5R200} is that the effect of \SigmaF~depends  
on galaxy stellar mass.
Within the factor of $\sim$13 between the low- and high-density bins,
\fred~increases by $\sim$ 0.2 for the intermediate-mass galaxies,
and $\sim$ 0.4 for the low-mass galaxies; whereas
for the high-mass galaxies, there is no statistically
significant difference in \fred.
The trends for the high- and intermediate-mass galaxies
appear to be very similar for galaxies in groups
of different \SMgrp, and also over the redshift range of
0.5 to 0.25. 
For the low-mass galaxies, the differentials in \fred~due to
different \SigmaF~may be somewhat larger for galaxies in massive groups. 
Thus, the magnitude of the local galaxy density effect
 appears to be fairly similar
for galaxies in groups of different richness across the
redshift range of 0.5 to 0.2, but has a strong dependence
on the stellar mass of the galaxies.

\subsubsection {Local Galaxy Density and Group Environment}
\begin{figure}
                \includegraphics[angle=90,width=8.2cm]{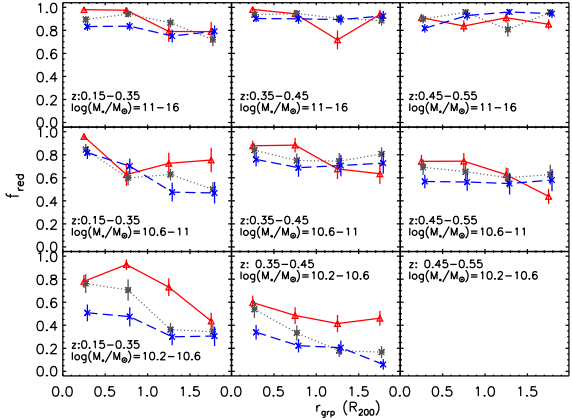}
                \caption{Red-galaxy fraction $f_{red}$ as a function of group-centric radius $r_{grp}$ for group galaxy subsamples of different local galaxy density $\Sigma_5$. The subsamples are further divided by redshift and group member galaxy stellar mass M$_*$ bins, as indicated on each of the 8 panels. The different $\Sigma_5$ subsamples from high to low are plotted in different linestyles as (red) solid, (gray) dotted, and (blue) long-dashed, respectively.
Groups of all total group mass M$_{*,grp}$ are combined for better signals.
                \label{dgcplot.r}}
                \end{figure}

The parameters \SigmaF~and \rgrp~are correlated.
To examine how these two effects interplay with each other, in
Figure \ref{dgcplot.r}~we present $f_{red}$ of samples of group galaxies of
different stellar masses as a function of $r_{grp}$ for
galaxies in three $\Sigma_5$ environments at different redshifts.
Since it appears that the group halo mass has only a weak 
influence on the effect of \SigmaF~on \fred, we combine groups
of different masses in order cut down the uncertainties in the 
\fred~measurements, which are still fairly large when the 
galaxy samples are further divided up into \rgrp~bins.

Inspection of Figure \ref{dgcplot.r}~shows that at all group-centric
radii, the local galaxy density has the strongest influence on the 
low-mass group galaxies, as is the case in Figure 13.
In general, the effect of \SigmaF~on \fred~is similar for galaxies at
different group-centric radii.
The most massive galaxies subsamples, again, all have high \fred~(larger
than 0.8) for all local densities and group-centric radii.
The \fred~of intermediate-mass galaxies shows a possible  weak change 
with local galaxy density with low significance, but consistent with that 
measured in Figure 13. 
For the low-stellar mass group  galaxies there is a clear
correlation between \fred~and local galaxy density.  
This is observed in all group-centric radii, including the core region.

Importantly, note that for the low-mass galaxies, even with local
galaxy density controlled, galaxies are on average redder in the center
of the group than those in the outskirts.
A similar, but weaker, trend is also seen for the intermediate-mass
galaxies.
This strongly suggests that the location of the (lower-mass) galaxies
in their parent group halo has an effect on their evolutionary history,
after the effects of local density and stellar mass are controlled.
That is, the redder population seen in the center of galaxy groups
and clusters is not due entirely to these galaxies being in a denser
region; being in the center of a massive halo also contributes to
them turning red.

\section{Discussion} \label{sec:discussion}
\subsection{The Uncertainties} \label{subsec:uncertainty}
Our work is exploited by photo-z, which is known to have less accurate redshift information than spectroscopical measurement.
With intricate corrections in weight and background subtraction, we discuss uncertainties in our procedure and possible effects on our results with the aid of mock catalog whose properties are detailed in \citet{2006MNRAS.365...11C}.
We assign photo-z to each mock galaxy such that the overall photo-z dispersion compared to mock redshift is comparable to that in our training set.

\subsubsection{$w_i$}
For each galaxy whose photo-z uncertainty is less than 0.6(1+$z$), we assign a weight $w_i$ based on the ratio of the number of all galaxies in the catalog to that in the sample in a given magnitude bin (\S2.2).
Such $w_i$ is only a function of magnitude, because we find that both blue and red galaxies, roughly separated by $B-R_c$=1.8, have comparable $w_i$.
The role of $w_i$ is to account for the missing galaxies in a given magnitude bin due to large photo-z errors.
In practice, for $n$ galaxies drawn from the sample, the expected total galaxy count is $\sum_{i=0,n} w_i$. 

To estimate the effect of $w_i$, we apply the same $w_i$ computation to the mock catalog with the simulated photo-z.
The $w_i$ reaches 2 at $R_c$=22.45, typical to our observed data.
We then compute $f_{red}$ for the mock galaxies at each redshift bin.
We find that such $f_{red}$ is smaller by $\sim$10\% than that using the known mock redshifts (i.e., no galaxy is removed hence no $w_i$ correction) at each redshift bin.
In other words, the $w_i$ correction results in a smaller $f_{red}$ but such effect is uniform over different redshift bins.
The zeropoint of the $f_{red}-z$ trend is smaller with $w_i$ applied, but the slope of the trend remains the same.

\subsubsection{Missing Groups and Contaminated pFoF Groups}
Our group sample is constructed using the pFoF algorithm on the photo-z catalog.
No group finding algorithms are known to be perfect, including the pFoF algorithm. 
We apply the pFoF algorithm to the mock catalog with the simulated photo-z and computed $w_i$.
We test the performance of the algorithm and investigate the roles of
 missing mock groups and falsely detected and contaminated FoF groups.

First, we find that the pFoF algorithm is able to recover $\geq$80\% of the mock groups whose halo mass are $\geq 2\times10^{13}M_{\sun}$ within our 
redshift range, and the recovery rate is not a strong function of redshift.
We compute $f_{red}$ using mock galaxies brighter than $M_{R_c}^*+1.5$ in each mock group, 
and investigate the average $f_{red}$ with and without any missing mock groups.
Since the recovery rate depends on halo mass, we carry out the comparison in three halo mass bins.
We find that for groups with halo mass greater than $10^{13}M_{\sun}$, 
approximately the lower limit of our sample, the missing groups have essentially
no effect on the average \fred, with the value being 1 to 2 \% redder for
the pFoF recovered groups when compared to the complete mock group catalog.
It is only by going down to the halo mass range
in halo mass to 1 to 7$\times10^{12} M_{\sun}$, well below our nominal
group mass limit, where we find an effect of about 10\%.

Second, we find the fraction of false and contaminated mock pFoF groups
is a strong function of redshift and $M_{*,grp}$.
We define a contaminated group as a pFoF group with a match from the
mock group catalog, but with a significant number of linked galaxies
not belonging to the true mock group (see Li \& Yee for a precise
definition).
In principle, in a large sample the effect of groups deemed contaminated 
on the measurement of \fred~is expected to be statistically corrected by 
the background corrections.
However, groups contaminated with larger fore- or background galaxies would
be more likely to be identified by any FoF algorithm. 
Thus, we include both false detected groups and contaminated groups in our
tests on their effects on the measurements of \fred.  Hereafter, we will
simply refer to these groups as `contaminated groups.'
 The contaminated groups tend to occur at higher redshift
 or small $M_{*,grp}$ regime.
Since galaxies in contaminated groups are expected to be bluer on average,
contaminated pFoF groups makes the $f_{red}$-$z$ trend slightly steeper 
in the low $M_{*,grp}$ bin.
For the low-mass pFoF mock groups with $M_{*,grp}$ 
between $10^{11.2}$ and $10^{11.8}M_{\sun}$, the contaminated group fraction 
is $\sim$20\% at $z\ltapr$0.45, but increases to $\sim 67$\% at $z\sim$0.5.
Using the mock groups we find $f_{red}$ for the mock pFoF sample at $z\sim$0.25
to be a factor of $\sim1.07$ smaller than the true value, with the discrepancy 
increasing to  $\sim1.23$ smaller at $z\sim$0.5.
For the more massive pFoF mock groups, the contaminated fraction is only
 $\sim$15\% at $z\sim$0.5, 
resulting in a smaller average $f_{red}$ by $\sim$3\%. 
We note that this small effect comes primarily from contaminated groups,
rather than false detection groups, as there are negligible number of
false massive groups.
The contamination fractions at lower redshift are smaller, resulting in
negligible effects on \fred~for the more massive groups.

We note that these pFoF mock groups are selected by $N_{gal}\geq5$, $N_{gz}\geq5$ and $M_{*,grp}$, but not by $B_{gc}$.
Based on Fig. \ref{bgc_mass}, the $B_{gc}\geq 125$ cutoff removes 10\% of the pFoF groups from the sample. 
Therefore, if they are further selected by $B_{gc}$, the above false detection
fraction and the difference in the average $f_{red}$ using the mock pFoF 
groups are expected to be even slightly lower.

As an example, In Figure \ref{fmstar.z} panel a), the contaminated pFoF groups 
in our sample make the $f_{red}-z$ trends steeper than the actual slopes. 
However, based on the above mock tests, we argue that 
the general declining $f_{red}-z$ trend cannot be simply caused by the increase
in contaminated pFoF groups, since the difference in $f_{red}$ between 
$z\sim$0.25 and $z\sim$0.5 is considerably larger than what is expected 
due to the effect of contaminated groups.
To account for the possible systematic effect of false group detection
and contaminated groups,
we increase the upper uncertainty error bars on \fred~for the low-mass
group bins based on our mock group tests in all Figures involving \fred.

\subsubsection{Group Centers and Group-centric Radius}
In \S3.3 we assign a pFoF linked galaxy to be the center of that galaxy group.
However, it is not necessary to have a galaxy located at the center.
We do so because this allows us to study properties of these `central' galaxies in the future.
Another way to define the group center is to take the median position of all the linked galaxies weighted by their luminosity and local galaxy surface density.
L09 find that such centers for the CNOC1 clusters are about 30\arcsec~from the cD galaxies.
For the RCS1 groups in this work, the median separation of these two types of group centers is $\sim$8\arcsec, which is equivalent to 0.04 Mpc at $z$=0.3 or 0.05Mpc at $z$=0.5.

The $R_{200}$ is another factor which affects the group-centric radius.
We compute the $R_{200}$ by extrapolating the $B_{gc}-R_{200}$ relation in \citet{2003ApJ...585..215Y}. 
The derived $R_{200}$ for our group sample is $\sim$1.07$\pm$0.24Mpc. 
Therefore, the uncertainty in $R_{200}$ dominates over the role of group center in computing $r_{grp}$.
We note that the $B_{gc}-R_{200}$ relation, which is derived based on the CNOC1 clusters, may not be the same for galaxy groups.
However, the relation indeed offers us an efficient way in estimating $R_{200}$ to be within an order of magnitude to the actual values.

The uncertainty in the $R_{200}$ introduces $\Delta log(M_{*,grp})\sim0.1M_{\sun}$ to the $M_{*,grp}$ in \S3.4, 
and gives a $\Delta f_{red}\sim0.04$ in the $f_{red}$ computed using group galaxies within 0.5$R_{200}$ of the groups.
We believe that our conclusions remain the same even though we have large uncertainties in group-centric radii, 
because our binsize for $M_{*,grp}$ is large enough to accommodate this extra uncertainty due to $R_{200}$, and any changes in $f_{red}$ are bigger than 0.04 in general.
\subsection{Properties of Group Galaxies}
  It has been known for decades that galaxy properties correlate with their 
environment: Galaxies in clusters are dominated by a red population, 
and field galaxies are characterized by a blue population. 
The morphology-density relation presented in \citet{1980ApJ...236..351D} has set the stage for many studies in the following decades to quantify this relationship and to investigate the causes. 
With the halo model of galaxies and groups, \citet{2006MNRAS.366....2W} suggest that the morphology-density relation can be expressed in terms of halo mass instead of galaxy number density, because the projected galaxy number in a halo is expected to correlate with the halo mass. 
A standard picture of the halo model is that some galaxies are embedded in their own individual halos independently, and some reside within a common halo 
shared with other galaxies in the form of galaxy groups. 
Once a halo enters within the virial radius of a more massive one, such as when a single-galaxy halo enters a group halo, it is referred to as a satellite halo or sub-halo within the larger halo (or, the parent halo). 
As the satellite or sub-halo orbits within the more massive one, it may
be subjected to all kinds of environmental effects;
 its mass is reduced and the diffuse outer part can be stripped off by tidal effects and interactions with other substructures. 
In addition to the mass loss, the gas reservoir of the infalling galaxy will be deprived and eventually lead to a halt in its star formation when the galaxy depletes its cold gas. 
This is usually referred to as `strangulation' \citep[e.g.,][]{1999ApJ...527...54B,2000ApJ...540..113B,2008MNRAS.387...79V,2008ApJ...672L.103K,2009MNRAS.400..937M}.
In this section we discuss the results from our galaxy group sample 
in the context of the halo model and `nature' versus `nurture' scenarios.

\subsubsection{Four Factors that Affect the Group Galaxy Population}
We have identified one intrinsic and three
global parameters that affect the 
evolution of the population of group galaxies as measured by the
red galaxy fraction, \fred.
The underlying predominant parameter is the mass of a galaxy's own
halo (measured via the galaxy's stellar mass).
The other three,
(1) the mass of the parent group halo into which the galaxy has fallen,
(2) the position of the galaxy in the parent halo, and
(3) the local galaxy density, 
produce further effects which evidently
accelerate the evolution of the galaxies to their final red color.
Here we summarize the observed effects of these four parameters on 
\fred~as shown in Figures \ref{fmstar.z}--\ref{dgcplot.r}.

\noindent
$\bullet$ The galaxy population as measured by \fred~is strongly
dependent on the galaxy stellar mass; i.e., its own halo mass.
The most massive galaxies (log(\smass)$\geq$11, or $M_{Rc}\ltapr M^*_{Rc}$+0.6
for red galaxies)
are already mostly red in all group environments by $z\sim0.5$.
The effects of the group environment and the local galaxy density,
summarized below, are
primarily seen in galaxies with log($M_*/M_{\sun}$)$<$11.

\noindent
$\bullet$ The mass of the parent group halo affects the galaxy population
such that the red galaxy fraction is larger for galaxies in more massive groups.
This effect is observed after controlling for the stellar mass
of the galaxies and it is stronger for lower-mass galaxies.

\noindent
$\bullet$ The red galaxy fraction has a  dependence on the
location of galaxies in their parent halo, as
parametrized by the group-centric radius, $r_{grp}$, normalized
by $R_{200}$.
The \fred~for intermediate- and low-mass galaxies increases
toward the group center, with the radial gradient being steeper
for lower-mass galaxies.
For low-mass galaxies, there are also some indications that 
the \fred~gradient is steeper for galaxies in more massive groups.

\noindent
$\bullet$  There is a degeneracy in the two parameters, group-centric
radius ($r_{grp}$) and local galaxy density ($\Sigma_5$),
that measure the environment of group galaxies.
By examining galaxy samples of intermediate and low stellar masses in
bins of local galaxy density, we find that:
 1) for galaxies in
the same local galaxy density bin, there is still a radial
gradient in $\fred$;  
and 2) conversely, for galaxies at the same
group-centric radius, those in higher local galaxy density
regions have larger \fred.
These results indicate that both the position of the galaxy in
the parent halo and the local galaxy density around each individual galaxy
have an effect on its  evolutionary history.

In the following subsections, we discuss in more detail the dependence of \fred~
on these various parameters, and how it fits into
the general context of galaxy evolution.

\subsubsection{The Dependence on Galaxy Stellar Mass}
The {\it apparent} strength of the dependence of the $f_{red}$
on environmental effects, whether it is the mass of the group halo, the position
the galaxies in the halo, or the local galaxy density, is a strong
function of the stellar mass of the galaxies.
The biggest effects are seen in the lowest mass galaxies; while
for the most massive galaxies, those with 
log($M_*/M_{\sun}$)$\gtapr$11$M_{\sun}$,
little discernible environmental effect is found.
Furthermore, for these most massive galaxies the $f_{red}$,
which is close to 1, has no apparent dependence on redshift 
(within our redshift range of  $z\ltapr0.5$) regardless 
of their group environment.
The uniformity of the slope, color zero-point, and
dispersion of the stacked red-sequence (Figure \ref{cmr}) for galaxies
brighter than $M_R^*$ also points to the conclusion that the most
massive galaxies in our sample 
have completed their evolution to a quiescent state by $z\sim0.5$,
suggesting that they mostly formed early and have little or no star
formation since  $z \sim1$.
Our conclusion that the stellar mass of a galaxy is a more fundamental 
factor in galaxy evolution echoes the recent spectroscopic results 
\citep[e.g.,][]{2010MNRAS.tmp.1856G, 2010A&A...524A...2C, 2010MNRAS.tmp.1805L,2010MNRAS.tmp.1657S}.
This illustrates that our photometric-redshift samples are able to give reliable conclusions while it provides a group sample size much larger than 
any from spectroscopic surveys at $z \geq 0.15$.

However, that the lowest-mass galaxies appear to be most affected by
the group and local density environments does not necessarily mean
that these environmental factors produce disproportionately
larger effects on low-mass galaxies, or that massive galaxies
are immune to them.  Rather, it is likely that massive galaxies
are equally affected by their environment, but the (secular) galaxy
down-sizing evolution effect has already turned most massive galaxies
red well before  $z\sim0.5$.
And once they are red, the `nurturing' influence can no longer
be discerned--whether a galaxy is far away from a parent halo,
or in a low galaxy density region, it is already red.
The same environmental effects likely have operated on these massive
galaxies earlier on, accelerating the process of them turning
red, to different degrees for galaxies in different environments.
In this interpretation, we should be able to observe similar
kinds of dependence of \fred~on group-centric radius, group
halo mass, and local galaxy density for galaxies of larger masses
at higher redshifts.
By the same token, for intermediate-mass galaxies at lower redshifts,
we should expect them to have a smaller apparent dependence of
their galaxy colors on group or local galaxy density environments,
because many of them would have already turned red by the present epoch.

\subsubsection{The Dependence on Group Halo Mass}
In the halo model, the infall of a galaxy in a subhalo into the parent group
halo is a significant event in the evolution of the stellar component
of the galaxy.
The larger halo can affect the star formation history of the infalling
galaxy via a number of mechanisms that act on a global scale within the parent
group, such as strangulation, ram-pressure stripping, and tidal 
disruption.
The dependence of the galaxy population on
the mass of their parent group or cluster can provide insights into
the  efficiency of the various environmental mechanisms
that affect star formation history, and their relative importance.
There is some evidence from redshift surveys in the local universe 
that galaxy population
properties have a dependence on the mass of the parent group halo,
in that galaxies in more massive halos tend to be in a more
evolved, or quiescent, state 
\citep[e.g.,][L09]{2006MNRAS.366....2W, 2009MNRAS.394.1131K, 2008MNRAS.387...79V}.
At higher redshifts, \citet{2010A&A...509A..40I}, using groups with
$z$ up to 1 cataloged
from the zCOSMOS survey, found that the galaxy blue
fraction is larger for groups of lower richness.
We find in our sample a similar dependence in general,
and we examine this dependence in more detail as functions of both
galaxy stellar mass and group-centric radius.

The dependence of the galaxy population on the parent group halo
mass is primarily seen in lower-mass galaxies.
Figure \ref{fmstar.z} also shows that the effect has a redshift dependence.
By $z\sim0.25$, the difference in \fred~for galaxies in
groups of different masses is seen mostly in low-mass galaxies;
whereas at $z\geq0.35$, both intermediate- and low-mass galaxies 
show a significant dependence of \fred~on \SMgrp.
Further subdividing the samples by the group-centric radius,
we find that the increase in \fred~for the intermediate- and
low-mass galaxies occurs over the range of group-centric radius
from the core to outside the virial radius.
However, in the core of rich groups galaxies of different masses
have a narrower range of \fred~values, compared to those in
poor groups; i.e., the dependence on group-centric radius is
stronger for galaxies in rich groups than poor groups.
We note that some caution should be used when considering the strength of 
this effect, in that it is more difficult to define the center of a poor
group, which may cause a smearing of the radial dependence.
 Weinmann et al. (2006) observed a similar dependence 
of the \fred-\rgrp~trend on group mass at 
$z\sim0$ using their SDSS group sample, with their lowest
mass groups (which are less massive than our low group mass bin)
showing almost no dependence of \fred~on \rgrp.  
Over all, this dependence on the group halo mass can be described,
to borrow a popular term from galaxy evolution, as some sort of `down-sizing'
effect, in that the galaxy population in richer groups appears to
be on average in a more advanced stage of their evolution. 

  Our data demonstrate the additional influence that group 
environment has on their constituent galaxies,
 but does the effectiveness of this influence depends on the mass of 
the parent halo?  That is to say, is this apparent `group down-sizing' effect 
due to the different effectiveness (or prevalence) of the mechanisms
that quench the star formation of the accreted member galaxies?
While this may be the case, the same effect can also be
produced by a simpler explanation: The larger \fred~in more massive
groups can simply be a result of a hierarchical build-up
of structures, such that at any given epoch the more massive 
groups/clusters have been in existence for a longer
time than less massive groups.
In a scenario where galaxy stellar mass is the primary determinant
of the time scale of galaxy evolution,
we can consider that group environment is acting  as an accelerator 
for the galaxy to reach its final red quiescent state
\citep[e.g.,][]{2010A&A...524A..76B,2011MNRAS.412.2303B}.
At a fixed redshift, if the time scale for the growth of the group from 
the accretion of `field' galaxies is longer than that for the truncation of star formation in the galaxies, then the infalling 
galaxies would turn red faster than the rate of accretion of
more field galaxies, creating a situation where groups that
have been in existence for a longer time would have a larger fraction 
of red galaxies.
Under this scenario, clusters or groups of a fixed halo mass
would expect to have an overall bluer population at higher redshift
than those at lower redshift, since, even if their accretion histories
are similar, their accretion time scale is compressed \citep{2009MNRAS.400..937M}.
This situation would naturally produce the Butcher-Oemler effect 
for a sample of fixed group/cluster mass over a range of redshift.

Observationally it is impossible to trace precisely 
the evolution of the progenitors of a given group along the redshift axis.
 Hierarchically, structures grow from smaller systems into larger ones; 
a rich group at lower redshift is likely to have a smaller mass in the past. 
In Figure \ref{fmstar.z}, we use fixed M$_{*,grp}$ cutoffs at different
 redshift bins without considering such mass evolution. 
If we could trace the progenitors, i.e., distinguish the ancestors, the 
apparent Butcher-Oemler effect and the `group down-sizing' effect is 
expected to be even stronger. 
For example, a group at $z\sim 0.25$ is expected to have a smaller mass
at $z\sim0.5$, and hence a lower M$_{*,grp}$. 
Since groups of smaller M$_{*,grp}$ at a fixed redshift have smaller $f_{red}$, the \it `$f_{red}$-z' \rm gradient for this group in our example will become even steeper, making the `group down-sizing' effect more significant.
\subsubsection{Environmental Influences: The Effect of Local Galaxy Density
and Group-Centric Radius}

The local galaxy density (\SigmaF) and group-centric 
(or cluster-centric) radius (\rgrp) are two parameters 
that are often used to  measure the environment of galaxies.
In fact there has long been a running discussion in the literature as to
which is more fundamental in determining the galaxy population properties
of cluster galaxies
\citep[e.g.,][]{1980ApJ...236..351D,1993ApJ...407..489W}.
While they are correlated with each other, since local galaxy density is
higher in the center relative to the outskirts, these two parameters
can provide indications of different physical mechanisms.
The group-centric radius can be considered as a parameter sensitive
to the global environment of the gravitational potential and gas
content of the parent group halo.
Thus, \rgrp~can be used to gauge the importance of mechanisms
such as ram-pressure striping, strangulation, and global tidal
effects from the group dark matter halo mass.
On the other hand, local galaxy density measures the more immediate
environment, and is more pertinent to mechanisms that are
dependent on the existence of neighboring galaxies, such as harassment
and galaxy-galaxy interactions and mergers.

In most studies of group or cluster galaxies, where the
galaxy samples are usually too small to separate into subsamples
of both radial and local-density bins, correlations of
galaxy population properties with one or the other parameter
become ambiguous.
Is the change in the galaxy population as a function of 
group-centric radius a reflection of the different local galaxy densities
at different radii?
Or conversely, is the correlation between local galaxy density
and galaxy population primarily a result of the position of the
galaxies within the larger parent halo in which they reside?
These questions can be answered by examining  the properties of
galaxies at a fixed group-centric radius as a function of local
galaxy density, or for galaxies in regions of a specific local
galaxy density in different group-centric radii.
A similar but more bimodal view  is taken by some investigators
in which galaxies are simply divided into ``central" galaxies and
``satellite" galaxies \citep[e.g.,][]{2008MNRAS.387...79V}.

 Figure \ref{dgcplot.r}, which plots the \fred~of group galaxies in different
local galaxy density bins as a function of group-centric radius,
 demonstrates the residual effect of local galaxy density 
after accounting  for the stellar mass of the group galaxies and their
positions within their parent groups.  
Again, the low-mass galaxies show the largest effect (showing up as
the separations between points of the same symbols in each panel of
Figure 14) due to local galaxy density,
 where over the factor of about 13 in $\SigmaF$,
\fred~is changed by roughly 0.2 to 0.5, with the difference being more
significant at lower redshift.
This change is comparable to the changes seen in \fred~over the \rgrp~range
from  outside \r200~to the core.
Similarly, for the intermediate- and high-mass galaxies, the difference 
in \fred~induced by the local galaxy density is also the same
order as that seen for
these galaxies between $\sim$\r200 and the group core: 
$\Delta f_{red}\leq 0.1$ for the massive galaxies, and about 0.1 to 0.2 for
the intermediate galaxies.

In most studies of the effect of local galaxy density on the
make-up of galaxy populations, the location of the galaxy within
a larger halo is rarely  considered (since often, these are ``field"
samples).
Furthermore, the stellar mass effect is clearly dominant over
that of local galaxy density.
Thus, if neither of these are controlled, the local galaxy density 
correlation could become misleadingly strong.
There are several investigations of the effect of local galaxy
densities for field galaxies in which the mass or luminosity
 of the galaxies is taken into consideration.
Using the same RCS1 photometric redshift catalog as a field sample
between $z$ of 0.2 and 0.6, \citet{2005ApJ...629L..77Y}
found almost no dependence of \fred~on local galaxy density at
$z\geq0.5$, and a weak dependence at $z\sim0.3$, whereas
\citet{2004MNRAS.348.1355B} found using the SDSS at $z\sim0$ a strong local
galaxy density dependence of \fred~for galaxies of all luminosities.
More recently, \citet{2006MNRAS.373..469B} and \citet{2010ApJ...721..193P}, using SDSS data, show that most of the dependence
of \fred~on environment primarily occurs for  galaxies
with log(\smass)$\leq 10.7$.

Our results, after controlling for group environment, show
a similar trend to that found by \citet{2006MNRAS.373..469B} for SDSS galaxies.
Because none of the three parameters, M$_*$, \SigmaF, and \fred,
are computed identically in different studies, we can only 
compare the dependence of \fred~from the two studies approximately.
Our three density bins center cover a range of approximately 1.1 dex in
\SigmaF, each with a half bin width of approximately 0.27 dex.
It is reasonable to match the local galaxy density parameters,
(\SigmaF~in this paper, and $\Sigma$~in \citet{2006MNRAS.373..469B}), by
assuming that the high ends of the density measurements are similar.
Hence, the high- and low-density bins in our work correspond
to approximately  log$\Sigma\sim0.8$, and -0.3 in Baldry et al.
>From their Figure 11, and averaged over similar bins of galaxy
stellar mass, the change in \fred, $\Delta$\fred, between local density bins equivalent to 
our low and high bins are $\sim0.1$, 0.2, and 0.3 for stellar masses
of log(\smass)$>$11.0, 10.6$\leq$log (\smass)$<$11.0, and
 10.2$\leq$log (\smass)$<$10.6, respectively.
These relative changes in \fred~for different local galaxy
densities  from the SDSS field galaxy sample
 are consistent with what we find for the group
galaxy sample at $z\sim0.25$, with $\Delta f_{red}\sim$0.1, 0.25, and 
0.4 for the three stellar mass bins, respectively.

We also find that the relative effect of local density on \fred~is
roughly constant as a function of group-centric radius and
the mass of the parent group halo.
This, along with the general agreement with the effect for field
galaxies, suggests that we can separate the influence between
local galaxy density and the global effects of the group
environment.
In the context of the debate of whether galaxy population is
correlated primarily with local galaxy density or
cluster/group-centric radius, our results indicate that both parameters
are important.

\subsubsection{Galaxy Groups as the `Nurturing' Influence}

In early discussions of
Dressler's work on the `morphology-density' relation (Dressler 1980),
along with the discovery of the Butcher-Oemler effect (Butcher \& Oelmer, 1978),
attempts to interpret the results along the lines of
the idea of `nature' versus `nurture' have often been used, though
not much has been resolved
\citep[e.g.,][]{1999ApJ...518..576P,2007MNRAS.376.1445C,2007MNRAS.374.1457M,2009A&A...503..379T}.
>From the perspective of the halo model, we can invoke this classic framework
in the following way.
The strong dependence of a galaxy's star formation 
history on its own halo mass can be attributed to `nature;' while
how it interacts with its immediate neighboring subhalos 
and how it is being influenced by the larger group halo into which 
it has been accreted can be thought of as the `nurturing' processes.
(There can be some debate as to whether the local galaxy density can
be considered as part of the `nature' process;
it could  be that a galaxy is more likely to be massive because it was 
formed in a high-density region.)
The dependence of the red galaxy fraction on the stellar mass 
of galaxies can be seen as a  manifestation of the
currently accepted  down-sizing star formation paradigm.
and considered as the underlying secular evolution of a galaxy.

After controlling
for both galaxy stellar mass and local galaxy density, our data clearly
show a residual dependence of the red galaxy fraction
on the group environment, especially for lower-mass galaxies.
Group galaxies show a \fred~gradient with the group-centric
radius, in that galaxies in the center are redder on average than
those further out.
Furthermore, this effect is larger in more massive groups.
This gradient  is not entirely due to the fact that local galaxy
density is higher in the core of a group, since it is still
seen using galaxy samples in fixed bins of $\Sigma_5$.
The change in \fred~going from outside $R_{200}$ to the core
is comparable to or larger than the changes seen in \fred~due to a change
of local density of a factor of $\sim13$ times.
This  suggests that the efficiency in truncating star formation
from physical processes arising from close neighbors is of
the same order as that due to the processes operating in the
larger halo into which the galaxy has been subsumed.

Physical processes that are related to neighboring galaxies
include galaxy mergers and harassment, producing strong
tidal effects on the galaxy.
We can consider the local galaxy density effect to be part
of the secular evolution, on top, or part, of the down-sizing effect;
i.e., these effects occur independently of whether a galaxy
with certain local galaxy density is sitting in the field, or in
the outskirts, or within the virialized region of a larger group.
However, the much more massive group halo may produce ram-pressure 
stripping due to intra-group gas, or a strangulation effect from 
the removal of the outer gaseous halo of the infalling galaxy. 
Such processes can serve as additional
mechanisms in driving the galaxy evolution towards the red-and-dead state.
The net effect is an acceleration of the truncation of 
star formation as the galaxy settles into the parent halo,
turning it red earlier than field galaxies of similar stellar
masses which are not affected by a group halo.

In the current literature, a number of investigators have separated
the effect of stellar mass and environment.
For example, \citet{2010ApJ...721..193P} modeled the evolution of galaxy
population using the concept of mutually independent ``mass-quenching"
and ``environment-quenching" efficiencies.
\citet{2009MNRAS.400..937M} used the idea of ``fraction of environmentally affected
galaxies", based on accretion history of galaxy groups and clusters
 from n-body simulations, to model the evolution of
cluster galaxy populations.
We note that these two studies effectively focused on two different
environmental parameters -- Peng et al.~on local galaxy density, and
McGee et al.~on group environment.
In the analysis of our large sample of group galaxies, we find
that we can in fact separate the environmental effect into
a local galaxy density effect and a more global large scale
structure effect due to the
presence of a larger group mass halo into which the subhalo has
been accreted.
These two types of effects likely invoke different mechanisms that
lead to environmental quenching of star formation in the
accreted galaxies, and both need to be accounted for explicitly
in modeling galaxy evolution.

In a picture where the
the dependence of star formation rate on stellar mass 
is considered as the long term secular trend in the evolution of a galaxy, 
the local galaxy density environment can accelerate the quenching rate of star
formation, but over different time scales, depending on the process.
For example, galaxy mergers can act as a one-time-only
process that happens stochastically and quickly change
 the nature of the galaxy; whereas
galaxy harassment can be a more gentle process acting over a longer
time scale.
The other major environmental event is
the accretion of a galaxy into a larger halo, which will further act
as an accelerator in the evolution of the galaxy.
This would be a short-time-scale, one-time-only event 
as the galaxy approaches the
group virial radius, where its star formation is quenched due to 
processes that are associated with the large gravitational
potential of the massive group halo.
A detailed study of dependence of star formation history 
of galaxies on these parameters based the mass of the galaxy, and
the local and global environments, will ultimately lead us to
a better understanding of galaxy evolution.

\section{Summary} \label{sec:summary}
Using 905 galaxy groups identified by the pFoF algorithm in the RCS1 
photometric-redshift sample and covering a group total halo mass range 
of $\sim10^{13.2}$ \msun~to $\sim 10^{14.5}$\msun,
we study the group galaxy population over
the redshift range  $0.15 \leq z < 0.52$. 
We examine the color-magnitude diagram of group galaxies as a function
of redshift and group richness, and
consider the effects of four parameters on the red galaxy fraction \fred~and 
its evolution: the galaxy stellar mass M$_*$, the total group stellar mass 
\SMgrp (as a proxy for the group halo mass), the group-centric radius \rgrp, 
and local galaxy density \SigmaF.

We find the bright end (brighter than $M^*_{Rc}$)
of the red-sequences in the CMD for stacked
groups in redshift and group halo mass bins to be remarkably uniform,
with their  zero point, slope, and dispersion consistent with those
found for clusters.
Thus, the bright end of the red sequence is already in place, and likely
formed at $z\gtapr2$, even for those in groups approximately
an order of magnitude less massive than clusters.

Most of the evolutionary effects are seen in galaxies of lower stellar mass
(strongest for M$_*\ltapr 10^{10.6}$\msun).
We find that groups at lower redshifts possess larger $f_{red}$ 
than those at higher redshifts, exhibiting a group Butcher-Oemler effect.
Examining the dependence of \fred~in more detail, we find:

\begin{enumerate}
\item
There is a strong dependence of \fred~on galaxy stellar mass.
More massive galaxies have larger \fred, and the group Butcher-Oemler
effect is seen within our redshift range only for galaxies 
with $M_*\ltapr 10^{11}$\msun.

\item
The strength of the dependence of \fred~on the 
environmental parameters is also a strong function of the galaxy stellar mass.
Galaxies with M$_*\gtapr 10^{11}$\msun~are almost all red, independent
of their local galaxy density, group-centric radius, and group halo mass.
In the items that follow, the results apply primarily for
galaxies with $M_*\ltapr 10^{11}$\msun.

\item
We find a dependence of \fred~on \SMgrp, in that galaxies in more massive
groups have a larger \fred.
This is seen after the group galaxies are separated into \rgrp~or \SigmaF~bins.
This effect is strongest for the lowest mass galaxies.
The group Butcher-Oemler effect appears to be stronger in lower-mass
groups over this redshift range; i.e., the change of \fred~from the
$z\sim0.5$ to the $z\sim0.2$ bin is larger for lower-mass groups.
This difference is probably at least in part due to low-mass groups
starting from a lower \fred~at $z\sim0.5$.

\item
There is a dependence of \fred~on group-centric radius, which is stronger
for lower-mass galaxies.
This dependence still exists after controlling for local galaxy density,
but is somewhat reduced.
The \fred--\rgrp~trend is weak for the low-mass groups, especially
in the two higher-redshift bins.
While this difference may be real, part of it may be contributed
by the larger uncertainty in determining the group center and/or
larger contamination for the lower-mass groups.

\item
While group-centric radius and local galaxy density are correlated,
nevertheless, at a fixed \rgrp, there is still a significant dependence
of \fred~on \SigmaF.
The typical change in \fred~over a factor of about 13 in \SigmaF~is
similar to that found in the lower-redshift SDSS ``field"-galaxy
sample, i.e., without controlling whether the galaxies are in clusters
or groups.
This change is also similar in magnitude to the change in \fred~seen
in galaxies from outside the virial radius to the group core.
The dependence on \SigmaF~is also larger for lower-mass galaxies.
This indicates that group environment has a residual effect over
that of local galaxy environment (or vice versa), and the two
must be considered at the same time.
\end{enumerate} 

A general picture for galaxy evolution, and in particular
for galaxies in groups and clusters, emerges from these 
correlations and other work in the literature at different redshifts.
Within the four parameters that we have examined,
in a `nature versus nurture' scenario, we can consider the galaxy stellar
mass as the predominant determinant of the evolutionary history
of a galaxy.
This would be the effect that produces the commonly accepted `down-sizing' of 
galaxy evolution.
The environmental influence on this secular trend, which accelerates
the galaxy to its final red state, can be considered as the `nurturing'
part of the galaxy's history.
There is some room left as to the debate of whether local galaxy density
is part of the intrinsic character of the galaxy, in that galaxies
born in a high density region may likely be  preferentially  more
massive.
However, local galaxy density, as an influence on galaxy evolution
may occur with different time scales and stochastically; e.g., galaxy 
harassment may be continuous, while galaxy mergers are a significantly
shorter time scale event that occur more rarely.
On the other hand, the effect of a galaxy falling into and being subsumed
by a more massive halo is most likely a one-time-only effect that
quenches the star formation in the galaxy over a relatively short time scale.
These environmental events all have lasting effects on a given galaxy
and the build-up of galaxy groups and clusters.
To truly understand galaxy evolution, we need to have 
samples that cover a wide redshift range and are sufficiently
large for us to clearly separate the effects of these parameters.
This will then allow us to
test the predictions, or inform the construction,
of a variety of models and simulations.
The recently completed RCS-2 \citep{2011AJ....141...94G} with a 
data set that will provide a similar sample of galaxy groups $\sim25$
times larger than the current sample will allow us to make great strides
toward this goal.

\acknowledgements
The data in this paper are 
based on observations obtained at the Canada-France-Hawaii Telescope (CFHT) which is operated by the National Research Council of Canada, the Institut National des Sciences de l'Univers of the Centre National de la Recherche Scientifique of France, and the University of Hawaii.
I.H.L acknowledges financial support from Swinburne University of Technology and the University of Toronto Fellowship. 
The RCS and the research of H.K.C.Y. are supported by grants from the Natural
Science and Engineering Research Council of Canada and the Canada Research
Chair program.  H.K.C.Y. also wishes to thank the Academia Sinica Institute of
Astronomy and Astrophysics for their hospitality during the latter stage
of the writing of the paper.


\end{document}